\documentclass[11pt,epsf]{article}
\DeclareMathAlphabet{\scr}{U}{rsfs}{m}{n}
\usepackage{latexsym}
\usepackage{epsfig}
\usepackage[mathscr]{eucal}
\usepackage{amsfonts}
\usepackage{amscd}
\usepackage{array}
\usepackage{amssymb}
\usepackage{colordvi}
\usepackage[centertags]{amsmath}
\usepackage{enumerate}
\usepackage{graphicx}
\usepackage{booktabs}
\usepackage{theorem}
\usepackage[footnotesize]{caption}
\usepackage{soul}
\usepackage{slashed}
\usepackage{textcomp} 
\usepackage[merge,numbers,compress]{natbib}

\usepackage[T1]{fontenc}
\usepackage[utf8]{inputenc} 

\usepackage{pdfpages}
\usepackage{url}
\usepackage{hyperref}

\pdfoutput=1

\newcommand{\sq}{\tilde{q}}
\newcommand{\sqbar}{\bar{\sq}}

\newcommand{\sul}{\tilde{u}_L}
\newcommand{\scl}{\tilde{c}_L}
\newcommand{\sdl}{\tilde{d}_L}
\newcommand{\ssl}{\tilde{s}_L}
\newcommand{\sur}{\tilde{u}_R}
\newcommand{\sdr}{\tilde{d}_R}
\newcommand{\ssr}{\tilde{s}_R}

\newcommand{\sbo}{\tilde{b}}
\newcommand{\neutone}{\tilde{\chi}^0_1}
\newcommand{\neuttwo}{\tilde{\chi}^0_2}
\newcommand{\neutthree}{\tilde{\chi}^0_3}
\newcommand{\neutfour}{\tilde{\chi}^0_4}
\newcommand{\charone}{\tilde{\chi}^{\pm}_1}
\newcommand{\chartwo}{\tilde{\chi}^{\pm}_2}

\newcommand\matB{{\cal B}}
\newcommand\matR{{\cal R}}

\newcommand\matO{{\cal O}}

\newcommand\matF{{\cal F}}

\newcommand{\as}{\alpha_s}
\newcommand{\HWG}{\textsc{Herwig++}}
\newcommand{\DS}{\textsc{Dipole-Shower}}
\newcommand{\PYTH}{\textsc{Pythia}}

\newcommand{\eq}[1]{\begin{equation} #1 \end{equation}}

\newcommand{\newc}{\newcommand}
\newc{\be}{\begin{equation}}
\newc{\ee}{\end{equation}}
\newc{\bi}{\begin{itemize}}
\newc{\ei}{\end{itemize}}
\newc{\benu}{\begin{enumerate}}
\newc{\eenu}{\end{enumerate}}
\newc{\bc}{\begin{center}}
\newc{\ec}{\end{center}}
\newc{\bfig}{\begin{figure}}
\newc{\efig}{\end{figure}}
\newc{\qbar}{\bar{q}}
\newc{\go}{\tilde{g}}
\newc{\PB}{\textsc{Powheg-Box}}

\newc{\dGa}{\text{d}\Gamma^{\sq_1 \rightarrow \neutone q}}
\newc{\dGb}{\text{d}\Gamma^{\sq_2 \rightarrow \neutone q}}
\newc{\Ga}{\Gamma_{\text{tot}}^{\sq_1}}
\newc{\Gb}{\Gamma_{\text{tot}}^{\sq_2}}
\newc{\GaLO}{\Gamma_{\text{tot},0}^{\sq_1}}
\newc{\GaNLO}{\Gamma_{\text{tot},1}^{\sq_1}}
\newc{\GbLO}{\Gamma_{\text{tot},0}^{\sq_2}}
\newc{\GbNLO}{\Gamma_{\text{tot},1}^{\sq_2}}
\newc{\ds}{\text{d}\sigma}
\begin{document}

\title{\hfill ~\\[-30mm]
\phantom{h} \hfill\mbox{\small KA--TP--19--2014}\\[-1.1cm]
\phantom{h} \hfill\mbox{\small SFB/CPP--14--57} \\[-1.1cm]
\phantom{h} \hfill\mbox{\small PSI--PR--14--06} \\[-1.1cm]
\phantom{h} \hfill\mbox{\small TTK--14--xx} 
\\[1cm]
\vspace{13mm}   \textbf{Squark Production and Decay matched with \\Parton Showers at NLO}}
\date{}
\author{
R.~Gavin$^{1\,}$\footnote{E-mail: \texttt{ryan.gavin@psi.ch}},
C.~Hangst$^{2\,}$\footnote{E-mail: \texttt{christian.hangst@kit.edu}},
M.~Kr\"amer$^{3\,}$\footnote{E-mail:
  \texttt{mkraemer@physik.rwth-aachen.de}},
M.~M\"uhlleitner$^{2\,}$\footnote{E-mail:
  \texttt{margarete.muehlleitner@kit.edu}}, \\
M.~Pellen$^{3\,}$\footnote{E-mail: \texttt{pellen@physik.rwth-aachen.de}},
E.~Popenda$^{1\,}$\footnote{E-mail: \texttt{eva.popenda@psi.ch}},
M.~Spira$^{1\,}$\footnote{E-mail: \texttt{michael.spira@psi.ch}}
\\[9mm]
{\small\it
$^1$Paul Scherrer Institut, CH-5232 Villigen PSI, Switzerland}\\[3mm]
{\small\it
$^2$Institute for Theoretical Physics, Karlsruhe Institute of Technology,} \\
{\small\it D-76128 Karlsruhe}\\[3mm]
{\small\it
$^3$Institute for Theoretical Particle Physics and Cosmology,}\\{\small \it RWTH Aachen University, D-52056 Aachen}\\
}

\maketitle

\begin{abstract}
\noindent  
Extending previous work on the predictions for the production of
supersymmetric (SUSY) particles at the LHC, we present the fully differential calculation of the next-to-leading
order (NLO) SUSY-QCD corrections to the production of squark and 
squark-antisquark pairs of the first two generations. The NLO cross
sections are combined with
the subsequent decay of the final state (anti)squarks into the
lightest neutralino and (anti)quark at NLO SUSY-QCD. No
assumptions on the squark masses are 
made, and the various subchannels are taken into account independently. 
In order to obtain realistic predictions for differential
distributions the fixed-order calculations
have to be combined with parton showers. 
Making use of the \textsc{Powheg} method we have implemented our results 
in the \textsc{Powheg-Box} framework and interfaced the NLO
calculation with the parton shower Monte Carlo programs 
\textsc{Pythia6} and \textsc{Herwig++}. The code is publicly available and can
be downloaded from the \textsc{Powheg-Box} webpage. The impact of the NLO
corrections on the differential distributions is studied and parton shower effects are
investigated for different benchmark scenarios.  
\end{abstract}
\thispagestyle{empty}
\vfill
\newpage
\setcounter{page}{1}

\tableofcontents
\newpage
\section{Introduction}
\label{ch:introduction}
Among the numerous extensions of the Standard Model (SM),
SUSY \cite{Volkov:1973ix,golfand,ramond,Wess,Wess2,Sohnius,Nilles,HaberKane,Gunion1,Gunion2,Gunion3} 
constitutes one of the most attractive and most
intensely studied options. SUSY allows to cure some of the flaws of the
SM like the hierarchy problem or the existence of Dark Matter, for
which SUSY with $R$-parity conservation provides a natural candidate. Thus one of 
the main tasks of the LHC is the search for SUSY particles. With the next
run of the LHC at high energy it will be possible to search for the
colour-charged SUSY particles, the squarks ($\tilde{q}$) and gluinos
($\tilde{g}$), in the multi-TeV
mass range \cite{CMS:2013xfa,ATLAS:2013hta,Gershtein:2013iqa}.  
In $R$-parity conserving SUSY, they are copiously produced in pairs
through the main SUSY-QCD production processes at the LHC, $pp \to
\tilde{q} \tilde{q}, \tilde{q} \overline{\tilde{q}}, \tilde{q}
\tilde{g}$ and $\tilde{g} \tilde{g}$.
 
The pair production cross sections for strongly-interacting SUSY
particles have been provided at leading order (LO) quite some time ago
\cite{squarklo1,squarklo2,squarklo3,squarklo4}. The NLO SUSY-QCD corrections have been 
completed about ten years later in
\cite{squarknlo1,squarknlo2,prospino,squarknlo3}. In these calculations the
squark masses have been assumed to be degenerate, with the exception
of stop pair production, where all squarks but the stop have been
taken degenerate. The NLO corrections turned out to be large,
increasing the cross sections by 5\% to 90\% depending on the process
and on the SUSY scenario under consideration. Furthermore, the inclusion of
the NLO corrections reduces the uncertainties due to the unknown
higher order corrections, reflected in the dependence of the cross
section on the unphysical factorization and renormalization scales,
from about $\pm 50$\% at LO to $\pm 15$\% at NLO.  In view of the
still large corrections at NLO, calculations have been performed
beyond NLO, including resummation and threshold effects
\cite{sqbnlo1,sqbnlo2,sqbnlo3,sqbnlo4,sqbnlo5,sqbnlo6,sqbnlo11,Borschensky:2014cia,sqbnlo7,sqbnlo8,sqbnlo9,sqbnlo10,sqthresh1,sqthresh2,sqthresh3}. These 
corrections lead to a further increase by up to 10\% of the inclusive
cross section and reduce the scale uncertainty further. Also electroweak
contributions have been considered \cite{ewlo1,ewlo2}, and their NLO corrections,
calculated in \cite{ewnlo1,ewnlo2,ewnlo3,ewnlo4,ewnlo5,ewnlo6,ewnlo7},
have been shown to be significant, depending on the 
model and the flavour and chirality of the final state squarks. 

The computation of the cross sections at LO and NLO SUSY-QCD can be
performed with the publicly available computer program
\textsc{Prospino} \cite{prospino_manual}. Based on the calculations in 
\cite{prospino,squarknlo3}, the NLO 
corrections, however, are only evaluated for degenerate squark
masses. Additionally, the loop-corrected cross sections for the
various subchannels of the different flavour and chirality
combinations are summed up. Though results for the individual
subchannels can be obtained, they are provided in the approximation of
scaling the exact LO cross section of the individual subchannel  with
a global $K$-factor, that is given by the ratio of the total NLO cross
section and the total LO cross section for degenerate squark
masses.\footnote{Note that this is only possible with the second version
  of \textsc{Prospino}, called \textsc{Prospino2}. Although the original
  version could be modified to return also results for the separate
  channels, in its public version it returns all LO and NLO subchannels summed
  up.} In this approximation it is assumed that the $K$-factors of
the different subchannels do not vary significantly. In principal, the program also
allows for the computation of the NLO differential distributions in the
transverse momentum and the rapidity of the SUSY particles, based on
the results in \cite{prospino}. There it was found that the distributions for
the investigated SUSY scenarios were only mildly distorted by the
NLO corrections, and it has thus been assumed that differential
$K$-factors are rather flat in general. 

Recently, results have been presented for the NLO SUSY-QCD corrections
to squark pair production without any simplifying assumptions on the SUSY particle
spectrum \cite{hollik,hollik2}, and including the subsequent NLO decays of
the final state squarks into a quark and neutralino.\footnote{A complete next-to-leading order study of
  top-squark pair production at the LHC, including QCD and EW
  corrections has been published in \cite{hollik3}.} 
In \cite{plehn} completely 
general NLO squark and gluino production cross sections based on the
\textsc{MADGOLEM} framework have been provided and compared to resummed
predictions from jet merging. In \cite{ownpaper,evathesis,owndiss}, we
have calculated the NLO corrections to the pair production of squarks
of the first two generations and implemented the cross section in a fully flexible partonic Monte
Carlo program without making any simplifying assumptions on the squark masses and 
treating the different subchannels individually. In the course 
of this calculation we have developed a new gauge-independent
approach for the subtraction of on-shell intermediate gluinos at the
fully differential level and
compared our approach to several methods proposed in the literature. Moreover, we
have extended the results \cite{hollik,hollik2,plehn}, by matching our
NLO calculation to parton showers using the \textsc{Powheg-Box}
\cite{nason,powheg,powhegbox} framework. 

These recent NLO calculations which take into account the full mass
spectrum have shown that the $K$-factors of the individual subchannels
can vary by up to 20\%. Therefore, in order to improve the accuracy of the cross
section predictions a proper NLO treatment of the individual
subchannels is necessary, without relying on an averaged
$K$-factor. Furthermore, it was found, that while the shapes of
semi-inclusive distributions are only mildly affected by NLO corrections, 
this is not the case for more exclusive observables. Here the $K$-factors 
can vary by up to $\pm 20$\% depending on the kinematics, both at the production 
level and after including squark decays supplemented by the clustering of
partons to form jets. Irrespective of the use of fixed or dynamical
scales, simply scaling LO distributions with a global $K$-factor
is not a good approximation for exclusive observables. 

In continuation of our effort to provide accurate predictions for SUSY
production processes at the LHC we present in this work our results
for the NLO SUSY-QCD corrections to squark-antisquark production of
the first two generations. We furthermore combine our results both for
squark pair production and for squark-antisquark production with the decay
of the (anti)squark into the lightest neutralino and (anti)quark at NLO
SUSY-QCD. All results are obtained at fully exclusive level and
without making any simplifying assumptions on the squark mass spectrum. In order
to obtain realistic predictions for exclusive observables we have combined our fixed-order NLO calculations 
with parton showers. To this end, the processes have been implemented
in the \textsc{Powheg-Box} framework \cite{owndiss,powhegbox} and
interfaced with different parton 
shower programs. The implementation has
been made publicly available and can be obtained from
\cite{powhegweb}. 

The outline of the paper is as follows. In section~\ref{ch:nlo} we
present the NLO calculation of the squark-antisquark
production process. Section~\ref{ch:decay} is devoted to the computation of the squark
decays at NLO and the combination with the production processes. Here
we study different approaches for the consistent combination at
NLO. The implementation in the \textsc{Powheg-Box} as well as our results at
fixed order and including parton shower effects are presented in
section~\ref{ch:res}. Finally, we compare our results with results
obtained with an approximate approach used in the SUSY searches by the
LHC experiments. We summarize and conclude in section~\ref{ch:conclusion}. 

\section{Squark-Antisquark Production at NLO}
\label{ch:nlo}
The calculation of the NLO corrections to squark-antisquark production is very similar to the one for squark pair production already presented in \cite{ownpaper}. Therefore, the following discussion summarizes only the main steps and points out the most important differences between the two processes. 
\subsection{Contributing Channels}
\begin{figure}
\centering
  \includegraphics[width=7.5cm]{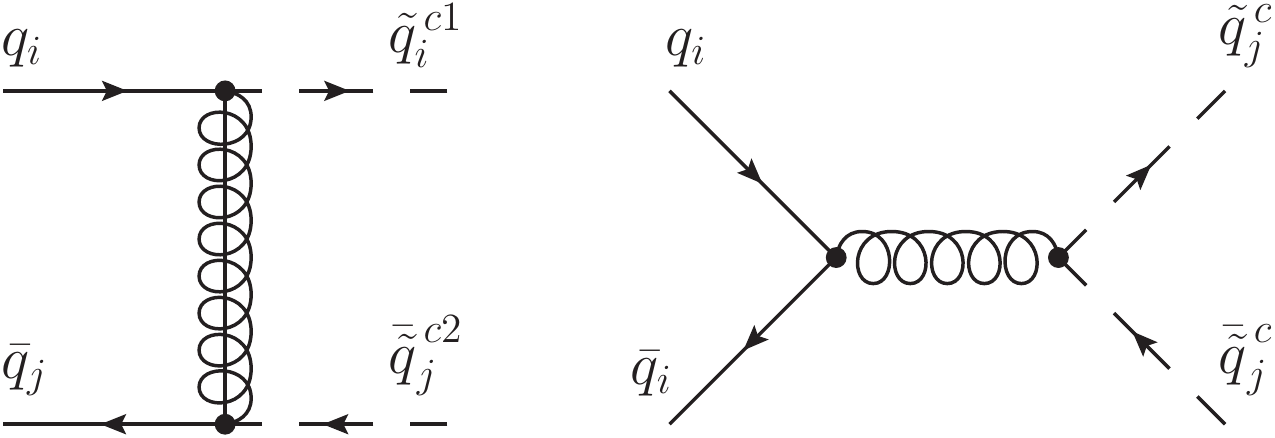}\\
  \vspace{0.5cm}
  \includegraphics[width=15cm]{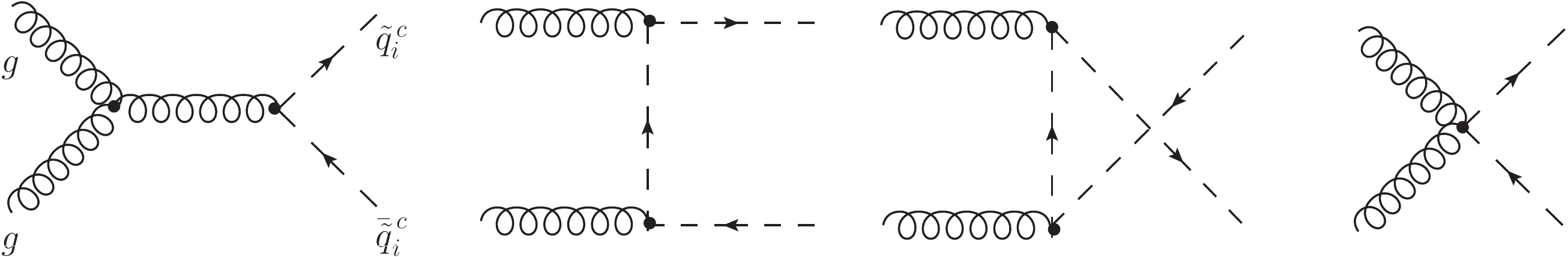}
  \caption{Feynman diagrams contributing to squark-antisquark production at LO.}
  \label{fig:sqsqbarLO}
\end{figure}

The production of a squark-antisquark pair at LO either proceeds via a pair of gluons or a quark-antiquark pair in the initial state: 
\begin{equation}
\begin{aligned}
 q_i\, \qbar_j &\rightarrow \sq_k^{\,c1}\, \sqbar_l^{\,c2}\, ,\\ 
g\, g &\rightarrow \sq_i^{\,c}\, \sqbar_i^{\,c}\, .
\end{aligned}
\label{eq:borncontri}
\end{equation}
Here, the lower indices indicate the flavour of the particle, whereas the upper indices for the squarks denote the respective chirality. The contributing Feynman diagrams are depicted in Fig.~\ref{fig:sqsqbarLO}. Due to the flavour conserving structure of the occurring vertices the $gg$ initiated diagrams and the $s$-channel diagram contribute only to the production of squarks of the same flavour and chirality. The results for the individual matrix elements squared can be found in \cite{owndiss}.

We consider in the following only the production of squarks of the first two generations mediated by the strong interaction. Correspondingly, the higher-order calculation comprises only SUSY-QCD corrections. In total, this leads to 64 possible final state combinations. This number can be reduced to 36 independent channels if the invariance under charge conjugation is taken into account. The number of independent channels can be reduced further if some of the squark masses are degenerate, as in this case the results for the $q\qbar$ initiated contributions differ only in the respective PDFs. However, we perform the calculation for a general mass spectrum and take advantage of this point only in the numerical analysis.

\subsection{Virtual and Real Corrections}
\label{sec:virtreal}
At NLO the squark-antisquark production processes receive contributions from virtual and real corrections. For the calculation of the virtual corrections we use the \textsc{Mathematica} packages \textsc{FeynArts 3.8} \cite{feynartsorig,feynarts,feynartsmssm} and \textsc{FormCalc 6.1} \cite{formcalc&looptools,formcalc2}. 
The numerical evaluation of the loop integrals is performed with \textsc{Looptools 2.7} \cite{formcalc&looptools,looptools2}.

In order to regularise the occurring ultraviolet (UV) divergences we apply Dimensional Regularisation (DR) \cite{dimreg,Wilson,WilsonFisher,Bollini,Ashmore}. The UV divergences are absorbed into the fields and parameters of the theory by introducing renormalization constants. For the renormalization of the strong coupling constant we use the $\overline{\text{MS}}$ scheme and decouple the heavy particles, {\it i.e.} the gluino, the top-quark and the squarks, from the running of the strong coupling constant $\as$. In the numerical analysis the 2-loop results for the determination of $\as$ at the scale of the process are used, hence we require the 1-loop decoupling coefficient, which can be found {\it e.g.} in \cite{prospino,collins,bernreuther,decoup}. Dimensional Regularisation violates SUSY explicitly by changing the number of degrees of freedom of the gluon field, inducing a mismatch between the gauge and the Yukawa couplings beyond LO. At NLO this effect can be cured by adding a finite SUSY restoring counterterm to the counterterm of the Yukawa coupling, see \cite{martinvaughn}. With these steps it is possible to use the five-flavour $\as^{(5),\overline{\text{MS}}}$ in the numerical analysis. 
The occurring fields and masses are renormalized using on-shell renormalization conditions. As the relevant counterterms are not included in the \textsc{FormCalc} version we use, they had to be implemented by hand in the MSSM model file.

\begin{figure}
\centering
  \includegraphics[width=15cm]{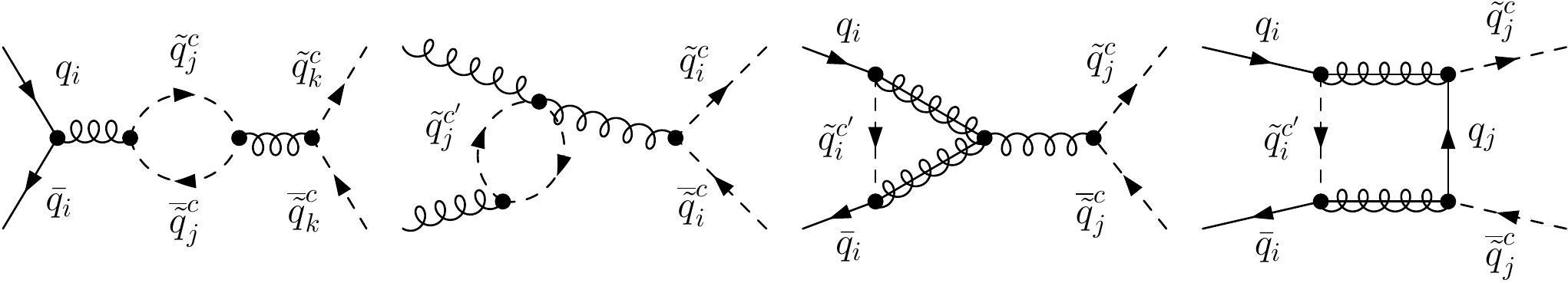}\\
  \vspace{0.5cm}
  \includegraphics[width=15cm]{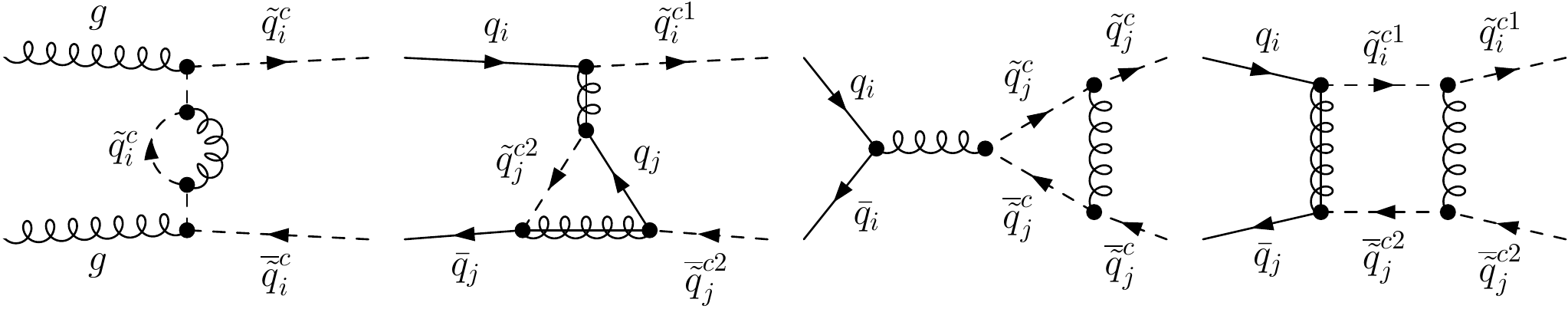}
  \caption{Sample Feynman diagrams contributing to the virtual corrections of squark-antisquark production.}
  \label{fig:sqsqbarvirt}
\end{figure}

The actual calculation of the corrections is performed such that the full mass dependence is preserved. In principle this requires the generation of all possible production modes with \textsc{FeynArts}, which is obviously a very inefficient procedure. Instead, we generated only the virtual contributions for $u \bar u\rightarrow \sul \bar{\tilde{u}}_L$, $u \bar u\rightarrow \sul \bar{\tilde{u}}_R$, $u \bar d\rightarrow \sul \bar{\tilde{d}}_L$, $d \bar d\rightarrow \sul \bar{\tilde{u}}_L$ and $ g g\rightarrow \sul \bar{\tilde{u}}_L$, where the indices $L$ and $R$ refer to the left- and right-handed chirality of the squarks. All other combinations of squarks in the final state can be traced back to one of these cases. However, this procedure requires a generalization of the masses of the internal squarks, if the corresponding propagators are connected to an external squark or quark line. In case of squark pair production this step amounted to simply replacing all internal squark masses in the vertex and box corrections with the masses of the external squarks, while the self-energy corrections could be left unchanged. For squark-antisquark production this generalization is more involved and requires a dedicated consideration of the individual diagrams. Some sample graphs are depicted in Fig.~\ref{fig:sqsqbarvirt}. The first two diagrams in the upper row are examples for the case where all internal masses have to be kept, {\it i.e.} here no changes are necessary. In the next two graphs the masses of the squarks in the loop have to be replaced case by case according to the flavour of the initial state quarks. Note that both chiralities have to be taken into account. The diagrams depicted in the lower row of the figure are examples for the case where one or more internal squarks are connected directly or indirectly to the final state squarks. The masses in the corresponding propagators and loop integrals have to be generalized accordingly.

The real corrections consist of the contributions with one additional gluon in the final state:
\begin{equation}
\begin{aligned}
q_i\, \qbar_j &\rightarrow \sq_k^{\,c1}\, \sqbar_l^{\,c2}\,g\, ,\\ 
g\, g &\rightarrow \sq_i^{\,c}\, \sqbar_i^{\,c}\,g\, .
\end{aligned}
\label{eq:realgcontri}
\end{equation}
Moreover, at NLO a new channel occurs with a gluon and an (anti)quark in the initial state:
\begin{equation}
\begin{aligned}
q_i\, g &\rightarrow \sq_k^{\,c1}\, \sqbar_l^{\,c2} q_j\, , \\
g \, \qbar_j &\rightarrow \sq_k^{\,c1}\, \sqbar_l^{\,c2} \qbar_i\, .
\end{aligned}
\label{eq:realcontri2}
\end{equation}
These channels are related to each other by invariance under charge conjugation.

In order to calculate the $q_i \qbar_j$, $q_i g$ and $g \qbar_j$ channels it is sufficient to perform the calculation for one of them and construct the other combinations by either crossing the gluon or by charge conjugating the respective process. Here, the calculation is performed analytically for the $q_i\, g \rightarrow \sq_k^{\,c1}\, \sqbar_l^{\,c2} q_j$ subprocesses. The occurring traces are evaluated with \textsc{FeynCalc 8.2} \cite{feyncalc}. The calculation is performed using two gauges for the external gluon, the Feynman gauge and a light-cone gauge. 

The $gg$-channels are obtained from \textsc{MadGraph 5.1.3.1} \cite{madgraph3} by generating the HELAS calls \cite{helas} for the specific process $g\, g \rightarrow \sul\,\bar{\tilde{u}}_L\,g$, generalizing the masses of the occurring squarks and removing the widths of the intermediate particles.

All these contributions exhibit infrared (IR) divergences, which cancel by virtue of the Kinoshita-Lee-Nauenberg theorem \cite{Ki62,LeNa64} against the corresponding divergences in the virtual contributions. As apt for a Monte Carlo event generator this cancellation is achieved by means of a subtraction formalism. We employ the FKS method \cite{fks}, which is automated in the \PB. 
  
\begin{figure}
\centering
  \includegraphics[width=15cm]{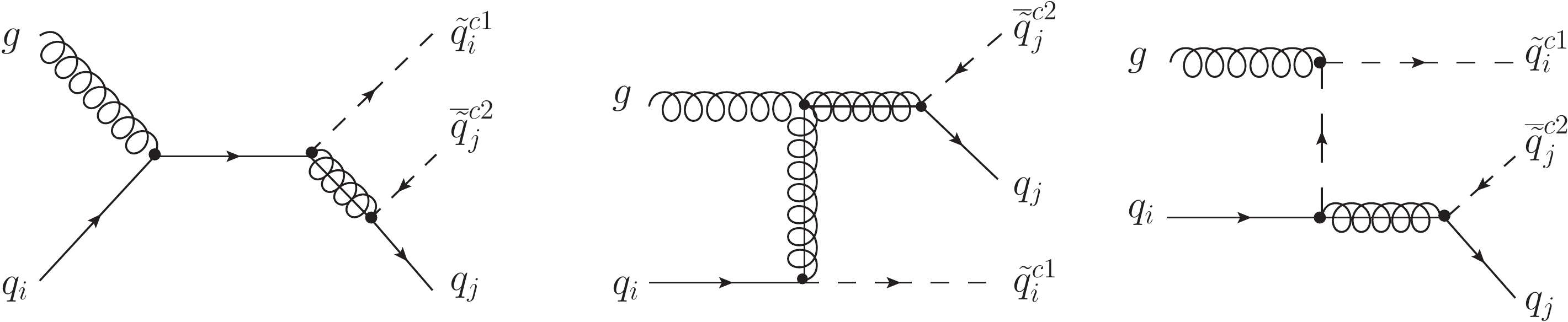}\\
  \caption{Feynman diagrams contributing to the real corrections of squark-antisquark production with potentially on-shell intermediate gluinos.}
  \label{fig:realqg}
\end{figure}

In the $qg$-initiated channels $q_i\, g \rightarrow \sq_i\, \sqbar_j q_j$ a second type of singularity occurs for scenarios with $m_{\go}>m_{\sq_j}$.\footnote{An equivalent problem appears in the $\qbar_i\, g \rightarrow \sq_j\, \sqbar_i \qbar_j$ channels. However, these contributions are related to the $q_i g$ case by charge conjugation and have been treated accordingly. They will not be discussed explicitly in the following.} For these mass configurations the intermediate gluino in the diagrams depicted in Fig.~\ref{fig:realqg} can be produced on-shell, causing a resonant behaviour. A similar problem has already been encountered in the calculation of squark pair production \cite{prospino,ownpaper}. Being formally equivalent to the Born contribution of on-shell squark-gluino production with the gluino decaying subsequently into a quark and an antisquark these contributions are large and require a proper definition of the process of interest. Keeping these terms would cause a double counting if the predictions for squark-antisquark production were combined with the ones for squark-gluino production. Hence, in order to obtain a meaningful result these on-shell contributions have to be subtracted consistently. 

There exist several methods to cope with this type of singularities, which have been developed in the context of $tW$ production \cite{twmcatnlo}, squark pair production \cite{ownpaper,hollik} and squark/gluino production \cite{prospino}. These approaches can be categorized as follows:
\bi
  \item \textbf{Diagram Removal (DR):} In this approach the resonant contributions are removed by either completely neglecting the Feynman diagrams in Fig.~\ref{fig:realqg} (DR-I) or by keeping the interference terms with the non-resonant contributions, but removing the amplitude squared of the depicted graphs (DR-II). Both approaches are rather easy to implement in a Monte Carlo program, but break gauge invariance.
  \item \textbf{Diagram Subtraction (DS):} These methods aim at a pointwise subtraction of the on-shell contributions by constructing a counterterm and performing a suitable reshuffling of the momenta. Hence both the interference terms and the off-shell contributions are kept by construction. In order to regularise the singular behaviour for $(p_{\sqbar_j}+p_{q_j})^2\rightarrow m_{\go}^2$ a finite width $\Gamma_{\go}$ for the resonant gluino has to be introduced (in fact this is also required in the DR-II scheme in order to regularise the integrable singularity in the interference terms). In the original proposal for $tW$ production \cite{twmcatnlo} this is achieved by replacing the corresponding propagator:
    \be
   \label{eq:bwprop}
     \frac{1}{(p_{\sqbar_j}+p_{q_j})^2-m_{\go}^2}\rightarrow\frac{1}{(p_{\sqbar_j}+p_{q_j})^2-m_{\go}^2+i m_{\go}\Gamma_{\go}}\quad.
    \ee
    However, this approach is only gauge invariant in the limit $\Gamma_{\go}\rightarrow 0$. A fully gauge invariant modification of the DS scheme (denoted DS$^*$ in the following) has been proposed in the context of squark pair production \cite{ownpaper}. In this approach the analytic expression for the amplitude squared is expanded in the poles $(p_{\sqbar_j}+p_{q_j})^2-m_{\go}^2\equiv s_{jg}$ before introducing the regularising width:
\be
   |M_{\text{tot}}|^2 = \frac{f_0}{s_{jg}^2}+\frac{f_1}{s_{jg}}+f_2(s_{jg}).
 \ee
 The coefficients $f_k$ ($k=0,1,2$) are gauge invariant quantities, i.e. introducing a regulator $\Gamma_{\go}$ at this point preserves gauge invariance and leads to
  \be
   \label{eq:expan}
   |M_{\text{tot}}|^2 = \frac{f_0}{s_{jg}^2+m_{\go}^2\Gamma_{\go}^2}+\frac{s_{jg}}{s_{jg}^2+m_{\go}^2\Gamma_{\go}^2} f_1 +f_2(s_{jg}).
 \ee
 The differences between the expressions obtained with the DS$^*$ and with the \lq usual' DS method vanish for $\Gamma_{\go}\rightarrow 0$ as expected, see \cite{ownpaper}. The counterterm for the subtraction of the on-shell contributions in this method is given by $f_0$ and reproduces the one used in the DS scheme in the limit $(p_{\sqbar_j}+p_{q_j})^2\rightarrow m_{\go}^2$. For more details on the momentum reshuffling and the construction of the subtraction term see \cite{ownpaper}.
\ei

 \begin{table}
\renewcommand{\arraystretch}{1.2}
\small
\bc
\begin{tabular}{|c || c |c  || r ||r | r|c|r| }\hline
Process  &  $\sigma^{\text{DS}^*} [\text{fb}]$  &  $\sigma^{\text{DR-II}} [\text{fb}]$  & $\Delta_{\sigma} [\%]$ & $\sigma^{\text{DS}^*}_{qg} [\text{fb}]$ & $\frac{\sigma^{\text{DS}^*}_{qg}}{\sigma^{\text{DS}^*}} [\%]$ & $\sigma^{\text{DR-II}}_{qg} [\text{fb}]$ & $\frac{\sigma^{\text{DR-II}}_{qg}}{\sigma^{\text{DR-II}}} [\%]$\\\hline\hline
$\tilde{u}_{L}\bar{\tilde{u}}_{L}$  & $  1.74\cdot 10^{-1} $ & $ 1.67\cdot 10^{-1} $ & $ 4.09 $ & $1.60\cdot 10^{-3}$ & $ 0.92 $ & $-5.46\cdot 10^{-3}$ & $ -3.27 $\\ 
$\tilde{u}_{R}\bar{\tilde{u}}_{R}$  & $  2.31\cdot 10^{-1} $ & $ 2.24\cdot 10^{-1} $ & $ 3.06 $ & $-5.71\cdot 10^{-4}$ & $ -0.25 $ & $-7.56\cdot 10^{-3}$ & $ -3.38 $\\ 
$\tilde{d}_{L}\bar{\tilde{d}}_{L}$  & $  1.15\cdot 10^{-1} $ & $ 1.13\cdot 10^{-1} $ & $ 2.02 $ & $-3.38\cdot 10^{-3}$ & $ -2.94 $ & $-5.67\cdot 10^{-3}$ & $ -5.03 $\\ 
$\tilde{d}_{R}\bar{\tilde{d}}_{R}$  & $  1.64\cdot 10^{-1} $ & $ 1.62\cdot 10^{-1} $ & $ 1.37 $ & $-6.02\cdot 10^{-3}$ & $ -3.66 $ & $-8.25\cdot 10^{-3}$ & $ -5.09 $\\ 
$\tilde{u}_{L}\bar{\tilde{u}}_{R}$  & $  6.94\cdot 10^{-1} $ & $ 6.79\cdot 10^{-1} $ & $ 2.12 $ & $-9.44\cdot 10^{-3}$ & $ -1.36 $ & $-2.40\cdot 10^{-2}$ & $ -3.54 $\\ 
$\tilde{d}_{L}\bar{\tilde{d}}_{R}$  & $  2.41\cdot 10^{-1} $ & $ 2.36\cdot 10^{-1} $ & $ 1.91 $ & $-3.41\cdot 10^{-3}$ & $ -1.42 $ & $-8.15\cdot 10^{-3}$ & $ -3.45 $\\ 
$\tilde{u}_{L}\bar{\tilde{d}}_{L}$  & $  8.42\cdot 10^{-2} $ & $ 7.49\cdot 10^{-2} $ & $ 11.1 $ & $7.80\cdot 10^{-3}$ & $ 9.27 $ & $-1.55\cdot 10^{-3}$ & $ -2.07 $\\ 
$\tilde{u}_{L}\bar{\tilde{d}}_{R}$  & $  4.92\cdot 10^{-1} $ & $ 4.83\cdot 10^{-1} $ & $ 1.88 $ & $-6.90\cdot 10^{-3}$ & $ -1.4 $ & $-1.60\cdot 10^{-2}$ & $ -3.3 $\\ 
$\tilde{u}_{R}\bar{\tilde{d}}_{L}$  & $  4.84\cdot 10^{-1} $ & $ 4.74\cdot 10^{-1} $ & $ 2.09 $ & $-6.03\cdot 10^{-3}$ & $ -1.25 $ & $-1.63\cdot 10^{-2}$ & $ -3.44 $\\ 
$\tilde{u}_{R}\bar{\tilde{d}}_{R}$  & $  1.09\cdot 10^{-1} $ & $ 1.00\cdot 10^{-1} $ & $ 8.33 $ & $7.47\cdot 10^{-3}$ & $ 6.83 $ & $-1.64\cdot 10^{-3}$ & $ -1.64 $\\ 
\hline\hline
Sum  & 2.79 & 2.71 & 2.72 & -0.0189 & -0.677 & -0.0946 & -3.49  \\\hline 
\end{tabular}
  \caption{\label{tab:qgxs}The NLO cross sections for squark-antisquark production of the first generation obtained for the CMSSM point $10.4.5$ applying the DS$^*$ scheme (second column) and the DR-II method (third column), with $\Delta_{\sigma}\equiv\left(\sigma^{\text{DS}^*}  -  \sigma^{\text{DR-II}}\right)/\sigma^{\text{DS}^*}$. The charge conjugate channels are combined. The last four columns contain the numerical values for the quantity $\sigma_{qg}$ as defined in the text and the respective contribution to the full NLO cross section, again for both the DS$^*$ and the DR-II method.}
\ec
\vspace*{-0.2cm}
\end{table}

The comparison of these different subtraction methods for squark pair production revealed for the scenario considered in \cite{ownpaper} only discrepancies in the total cross section at the per mille level. Repeating this study for squark-antisquark production, however, leads to larger differences, as the contributions of the $qg$ initiated channels are larger in this case. To illustrate this point the predictions for the total production cross sections of squarks of the first generation as obtained with the DR-II (using the light-cone gauge) and the DS$^*$ scheme are summarized in the second and third column of Tab.~\ref{tab:qgxs}. The scenario considered here corresponds to the mSUGRA point $10.4.5$ \cite{susybench} specified in Sec.~\ref{ch:res}. For the regularising width we choose $\Gamma_{\go}= 1\,\text{GeV}$. As can be inferred from the percental difference between the respective numbers given in the fourth column the predictions obtained with these two methods differ by up to 11\%, leading to a discrepancy of 2.7\% after summing these channels. Taking into account the contributions of the squarks of the second generation, too, increases this discrepancy further:
\be
 \sigma^{\text{DS}^*} = 4.37\,\text{fb} \quad \textnormal{and} \quad \sigma^{\text{DR-II}}= 4.21\,\text{fb} \,,
\ee
corresponding to a discrepancy of $3.6\%$. 

The fifth and seventh column of Tab.~\ref{tab:qgxs} contain the respective predictions $\sigma_{qg}$ for the $qg$ contribution to each channel. This (unphysical) quantity comprises the $2\rightarrow 3$ parts of the respective channel, {\it i.e.} the real amplitudes squared and the corresponding FKS counterterm and hence allows for a direct estimation of the effects of the applied subtraction scheme. As can be inferred from the table these contributions make up several percent of the individual cross sections. Hence the large discrepancies observed between the two subtraction methods have significant effects on the predictions for the total cross sections.
 
\begin{figure}
  \begin{minipage}{0.49\textwidth}
 \includegraphics[width=\textwidth]{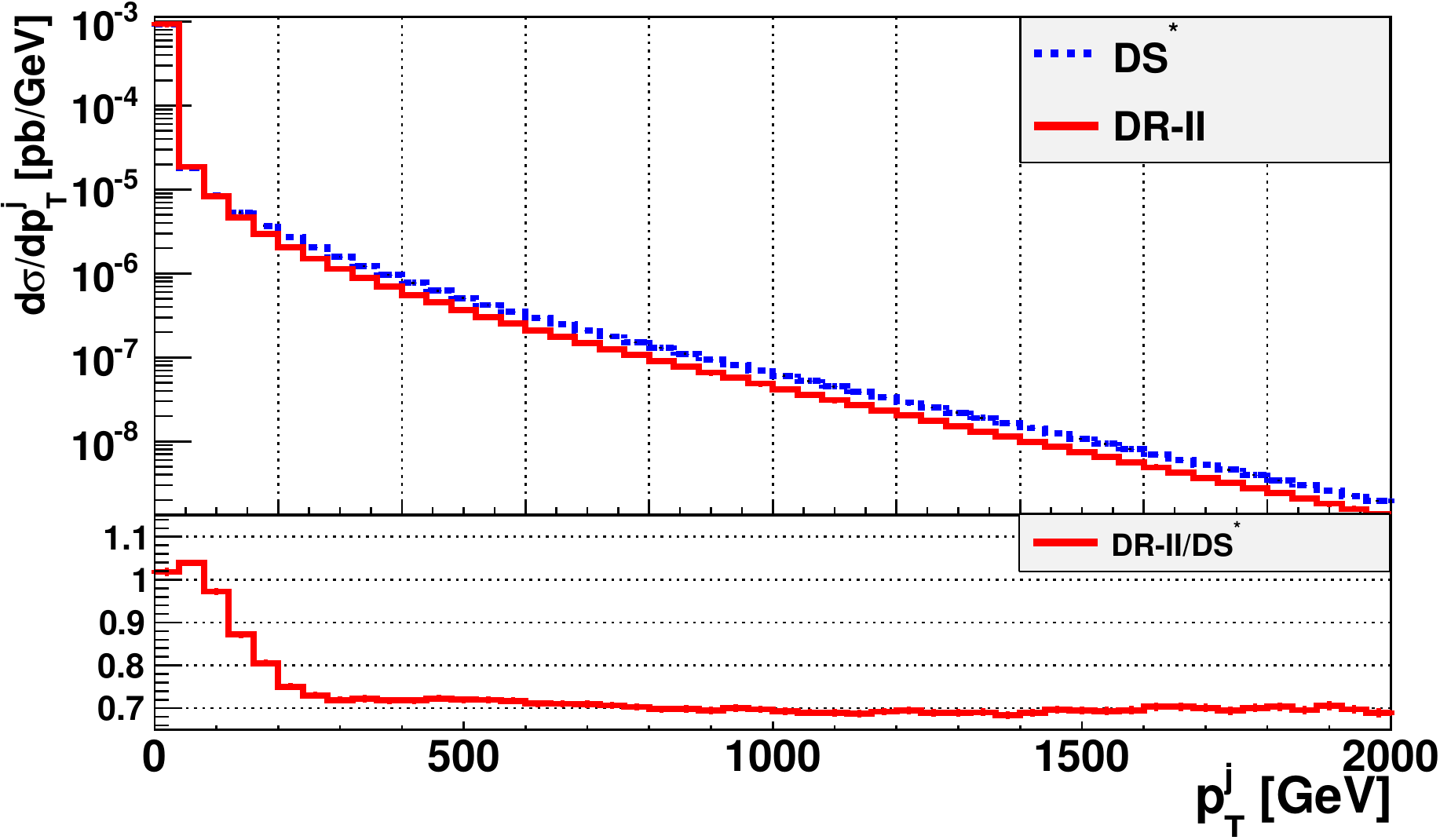}
\end{minipage}
\begin{minipage}{0.49\textwidth}
 \includegraphics[width=\textwidth]{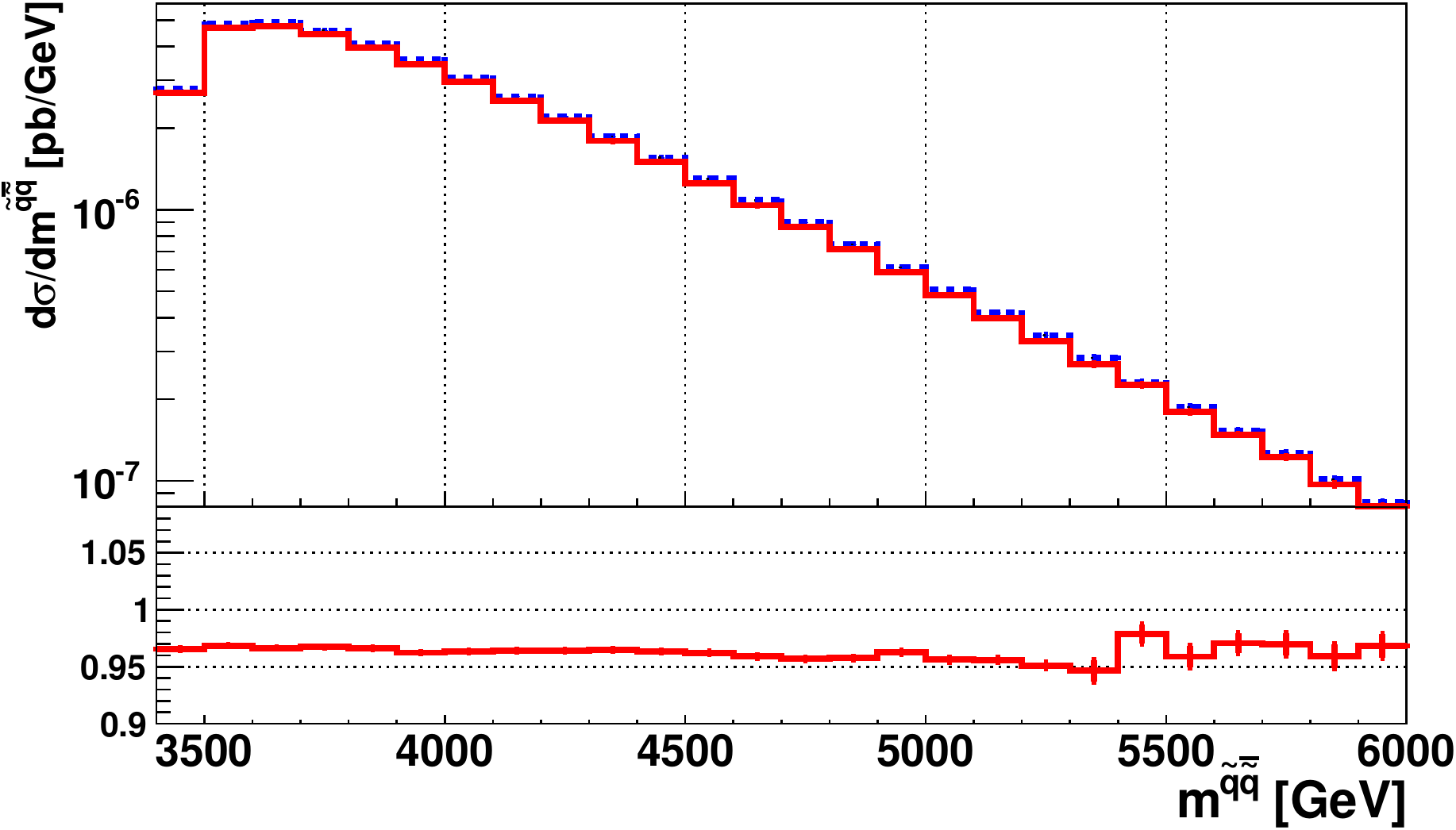}
\end{minipage}
\caption {The distributions for squark-antisquark production of the transverse momentum of the radiated parton generated in the real contributions, $p_T^j$, (left) and the invariant mass $m^{\sq\sqbar}$ (right) for the subtraction methods DS$^*$ and DR-II. The lower panels show the respective ratio of the DR-II and the DS$^*$ result.}
\label{fig:oss_sqantisq_distri}
\end{figure}

 Even larger effects of the chosen subtraction scheme can be observed in differential distributions which are sensitive to the emitted parton of the real corrections.
 As an example, the $p_T$ distribution of the radiated parton obtained with the DR-II scheme and the DS$^*$ method is shown in Fig.~\ref{fig:oss_sqantisq_distri} (left). For $p_T^j>200\,\text{GeV}$ the two predictions differ by about 30\%. In contrast, the shape of the $m^{\sq\sqbar}$ distribution (right plot in Fig.~\ref{fig:oss_sqantisq_distri}), which is supposed to be less sensitive to additional radiation, is not affected by the chosen method. Solely the normalization reflects the 3.6\% discrepancy already encountered in the total cross section

 \subsection{Tests and Comparison}
 The calculation presented in the last section has undergone numerous checks and comparisons. An obvious test for the correctness of the calculation consists in a comparison with the public program \textsc{Prospino2} for the limit of a mass degenerate spectrum. Unfortunately, a direct comparison of the results obtained with this public code is not straightforward, as it implicitly takes into account the sbottom production processes $g g \rightarrow \sbo \bar{\sbo}$ and $q \qbar \rightarrow \sbo \bar{\sbo}$, while the contributions $b \bar{b}\rightarrow \sbo \bar{\sbo}$ are neglected. Moreover, at NLO the contributions $q g \rightarrow \sq \bar{\sbo} b$ and the charge conjugate processes are taken into account. Instead of mimicking the way the total $K$-factor is calculated in \textsc{Prospino2} we have compared the numerical results of our calculation with a non-public implementation of the original results from \cite{prospino}, denoted \textsc{Prospino$^*$} in the following. Besides testing our calculation for the special case of degenerate squarks we have intensively checked the individual building blocks:
 \bi
 \item  The Born expressions have been compared with results given in the literature \cite{squarklo2,squarklo3,squarklo4}. In addition, the numerical comparison of the total cross section with \textsc{Prospino$^*$} provides a simple cross check for the correctness of the nontrivial combinatorics of the contributing channels.
 \item The UV finiteness of the virtual corrections has been checked both analytically and numerically. The correct structure of the IR poles has been verified by comparison with the known structure for the case of massive coloured particles in the final state, see {\it e.g.} \cite{madfks}. The correctness of the modifications performed in the virtual routines in order to generalize them to an arbitrary mass spectrum has been tested by performing this generalization for both $g g \rightarrow \sul \bar{\tilde{u}}_L$ and  $g g \rightarrow \sdl \bar{\tilde{d}}_L$ and comparing the outcome numerically. Likewise, the other cases mentioned in Sec.~\ref{sec:virtreal} have been checked.
 \item The analytic results for the real matrix elements squared have been compared numerically for a multitude of arbitrary phase space points with the routines generated with \textsc{MadGraph 5}. The cancellation of the IR poles against the FKS counterterms has been tested using the automatic procedure provided by the \PB. The gauge invariance of the DS$^*$ scheme has been explicitly checked by comparing the outcome of the two different gauges used in the calculation. Furthermore, the equivalence of the DS and the DS$^*$ scheme in the limit $\Gamma_{\go}\rightarrow 0$ has been verified numerically.
 \item The individual results for the three production channels $gg$, $q\qbar$ and $qg$ have been compared for degenerate mass spectra with \textsc{Prospino$^*$}.
 \ei

\section{Squark Decays at NLO and Combination with Production Processes}
\label{ch:decay}
The calculation of NLO SUSY-QCD corrections to production processes is only a first step towards a realistic prediction of possible events at the LHC. The next step requires the inclusion of the decays of the produced particles. Here, only the decay mode into a quark and the lightest neutralino for squarks, $\sq \to q \neutone$, or into an antiquark and the lightest neutralino for antisquarks, $\bar{\sq} \to \bar{q} + \neutone$, will be taken into account. In many SUSY scenarios, in particular the ones studied in this paper, the lightest neutralino is the lightest supersymmetric particle and stable (if R-parity conservation is assumed). The SUSY-QCD corrections to this decay have been known for several years. However, in the original calculation \cite{decay1} only results for the partial width have been given, so that a differential description is not possible. Recently, a fully differential calculation in the context of squark pair production and decay has been presented \cite{hollik} where the radiative corrections to the decay have been included by using the phase-space-slicing technique. In the next subsection we present a recalculation of the decay at the fully differential level by applying the subtraction method developed for single top production and decay \cite{proddec2} to our process. The second part of this section deals with the consistent calculation of the total squark width, which is required for the combination of the production and decay processes described in the last part of this section.

\subsection{Decay Width for $\sq \to q + \neutone$ at NLO}
\label{mt:decay}

\begin{figure}[!t]
 \centering
  \includegraphics[width=12cm]{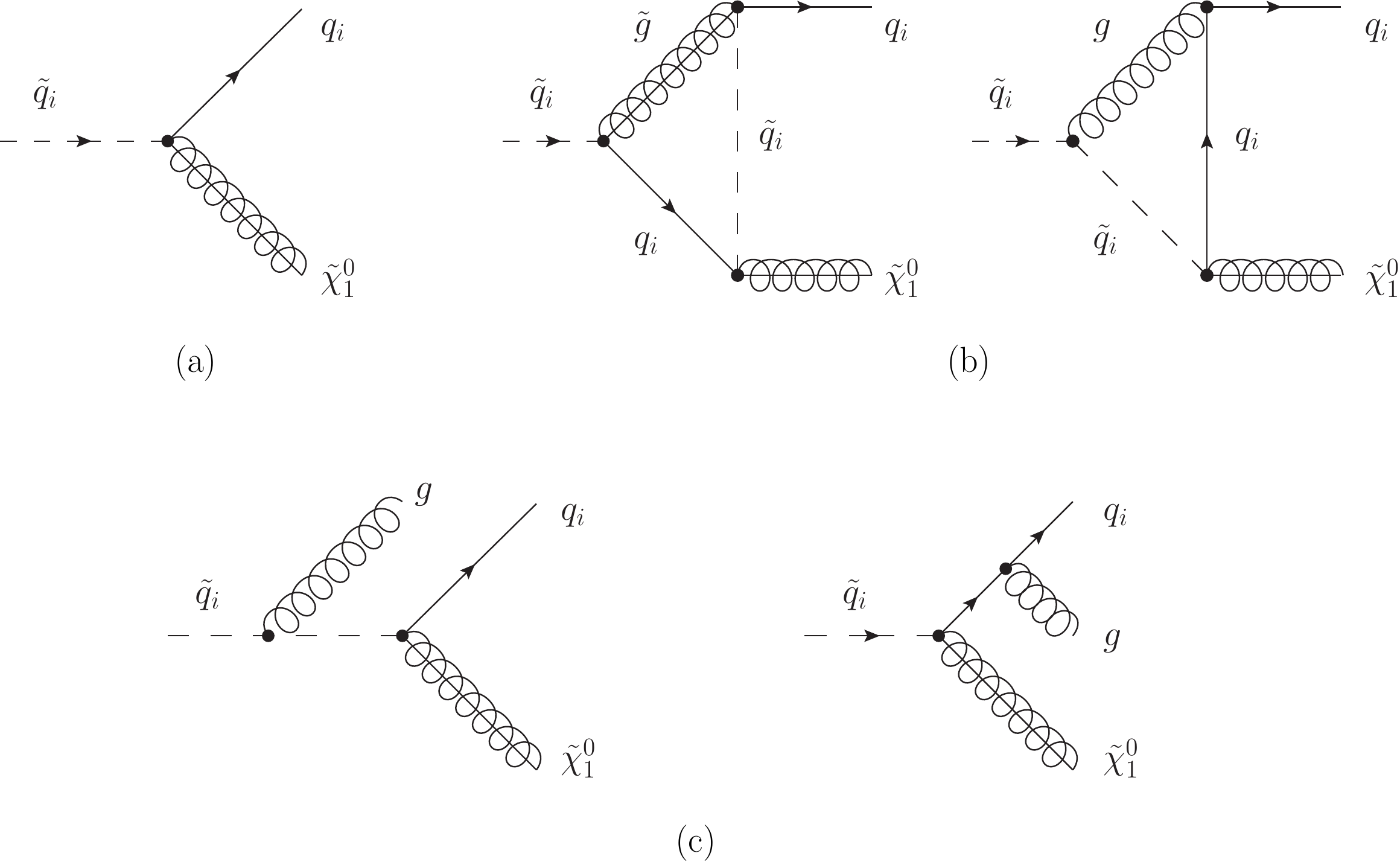}
  \caption{Feynman diagrams contributing to the decay $\sq_i\rightarrow q_i \neutone$ at LO  (a) and at NLO: virtual corrections (b) and real gluon radiation (c).}
  \label{fig:decgraphs}
\end{figure}
The LO contribution to the decay width of the process
\begin{equation}
 \sq \to q + \neutone
\end{equation}
comprises only one Feynman diagram, which is depicted in Fig.~\ref{fig:decgraphs} (a). 
As in the production process, at NLO virtual and real corrections have to be taken into account. The two diagrams contributing to the virtual corrections are shown in Fig.~\ref{fig:decgraphs} (b). The calculation is performed in \textsc{DR} and the external fields are renormalized on-shell. All integrals have been evaluated analytically. The package \textsc{HypExp} \cite{HuMa05, HuMa07} has been used to expand hypergeometric functions. Using \textsc{DR} requires, again, the introduction of a finite counterterm. Here, the squark-quark-neutralino Yukawa coupling $\hat{g}$ is affected. The SUSY-restoring counterterm leads to the following relation to the gauge coupling $g$ \cite{martinvaughn},
\begin{equation}
 \hat{g} = g \left[1-\frac{\alpha_s}{6 \pi} \right] \ .
\end{equation}
The real corrections involve an additional gluon,
\begin{equation}
 \sq \to q + \neutone + g  \ ,
\end{equation}
emitted either from the squark in the initial state or the quark in the final state, as displayed in Fig.~\ref{fig:decgraphs} (c). The IR divergences arising from the soft and/or collinear emission of the gluon cancel against the corresponding ones in the virtual corrections. This cancellation is achieved on the differential level by applying the subtraction method developed in \cite{proddec2} for single top production and decay to our decay process. The divergences in the real radiation process are cancelled by a local counterterm, which is constructed such that it has the same singular behaviour as the full matrix element. It takes the form of the LO matrix element squared multiplied by a function $D$, which describes the emission of the soft or collinear radiation:
\begin{equation}
 |\mathcal{M}_r(p_{\tilde{q}_i},p_q,p_{\tilde{\chi}^0_1},p_g)|^2 \rightarrow |\mathcal{M}_0(p_{\tilde{q}_i},p'_q,p'_{\tilde{\chi}^0_1})|^2 \times D(p_g \cdot p_{\tilde{q}_i},p_g \cdot p_q,m^2_{{\tilde{q}_i}},m^2_{\tilde{\chi}^0_1}) \ .
  \label{eq:sing}
\end{equation}
In the limit of soft emission, when $p_g \to 0$, or where the momenta of the quark $p_q$ and the gluon $p_g$ are collinear, the counterterm on the right-hand side of Eq.~(\ref{eq:sing}) has the same singular structure as the full matrix element squared on the left-hand side. The LO matrix element $\mathcal{M}_0$ is evaluated with modified momenta $p'_q$ and $p'_{\tilde{\chi}^0_1}$ which absorb the momentum carried away by the gluon. They are subject to momentum conservation $p_{\tilde{q}_i} = p'_q + p'_{\tilde{\chi}^0_1}$, as well as to the on-shell conditions $p'^2_q = 0$ and $p'^2_{\tilde{\chi}^0_1} = m^2_{\tilde{\chi}^0_1}$.

In the process at hand advantage can be taken from the fact that the LO matrix element squared can be easily factorized from the divergent part of the real matrix element squared,
\eq{|\mathcal{M}_r^\textnormal{Div}|^2 = \frac{4}{3}\ \frac{16 \pi}{m_{\tilde{q}_i}^2}\ \alpha_s\ |\mathcal{M}_0|^2 f(y,z) }
with the function $f(y,z)$, calculated in $d=4-2\epsilon$ dimensions, defined as
\eq{f(y,z) = - \frac12 \frac{1}{(1-\sqrt{r})^2} \left( \frac{1+z}{y} + \frac{1-z}{y} \ \epsilon \right) + \frac{1}{(1-\sqrt{r})^2} \frac{1}{y(1-z)} - \frac{1}{(1-r)^2 (1-z)^2} \; .}
In this function the following substitutions have been made
\eq{p_q \cdot p_g = \frac{{m_{{\tilde{q}_i}}}^2}{2} (1-\sqrt{r})^2 y \quad \textnormal{and} \quad p_{\sq} \cdot p_g = \frac{{m_{{\tilde{q}_i}}}^2}{2} (1-r) (1-z) }
with $r$ denoting the squared ratio of the neutralino over the squark mass,
\begin{equation}
 r = {m^2_{\tilde{\chi}^0_1}} / {m^2_{\tilde{q}_i}} \ . 
\end{equation}
The coefficient of the LO matrix element squared can then be chosen to serve as the divergent part, denoted by $D$, in the counterterm in Eq.~(\ref{eq:sing}),
\begin{equation}
 D(p_g \cdot p_{\tilde{q}_i},p_g \cdot p_q,m^2_{{\tilde{q}_i}},m^2_{\tilde{\chi}^0_1}) =\frac{4}{3}\ \frac{16 \pi}{m_{\tilde{q}_i}^2}\ \alpha_s\ f(y,z) .
\end{equation}
In order to cancel the IR divergences in the virtual corrections this counterterm needs to be integrated analytically over the one-particle phase space of the emitted gluon. The results for the necessary integrals can be found in Table 1 of \cite{proddec2}. The integrated counterterm then reads
\eq{\label{eqn6} \int d \Phi_1\ |\mathcal{M}_r^\textnormal{Div}|^2 = \frac{4}{3}\ \frac{\alpha_s}{\pi}\ \left(\frac{4\pi}{m_{\tilde{q}_i}^2} \right)^{\epsilon} |\mathcal{M}_0|^2 <f(y,z)(1-\sqrt{r})^2> }
with
\begin{eqnarray}
\hspace*{-0.9cm}
<f(y,z)(1-\sqrt{r})^2> &=& \frac{1}{2\,\epsilon^2} + \frac{5}{4\,\epsilon} - \frac1\epsilon \ln(1-r) - \frac52 \ln(1-r) + \frac{(7-5r)}{8(1-r)} -\mathrm{Li}_2(r) - \frac{7 \pi^2}{24}\nonumber \\
&-& \frac32 \frac{r}{1-r} \ln r + \frac14 \frac{r^2}{(1-r)^2} \ln r - \ln(r)\ln(1-r) + \ln^2(1-r) + \frac{11}{4} \; .
\label{eqn7}  
\end{eqnarray}
All steps of the analytical calculation have been checked against \cite{LaJe}. The results for the partial widths at LO and NLO have been compared to the result obtained from \textsc{Sdecay 1.3} \cite{sdecay}. Moreover, we have compared our results to the results presented in the independent calculation of squark pair production and decay of \cite{hollik}, in particular to the results given in Table 6 for the benchmark point 10.1.5 and the corresponding distributions, and have found agreement. In addition, this decay has been implemented in the \textsc{Powheg-Box}. The virtual corrections for this independent calculation have been calculated with \textsc{FeynArts/FormCalc} and the loop integrals have been evaluated with \textsc{LoopTools}. The real matrix elements squared, calculated by hand, have been tested numerically for a multitude of phase space points against the corresponding routines obtained with \textsc{MadGraph}. In the \textsc{Powheg-Box} the cancellation of the divergences is achieved automatically via the implemented FKS subtraction method. We have found perfect agreement between the calculation presented in this section and the implementation in the \textsc{Powheg-Box}.

\subsection{Total Squark Width at NLO}
\label{sec:totwidth}

For the calculation of the squark branching ratios we also need the total decay width $\Gamma^{\sq}_{\textnormal{tot}}$, both at LO and NLO. Furthermore, the NLO total decay width will be necessary to normalize the expressions for the combination of the production and decay processes, as we will see in the next subsection.
Since we only consider the decay into a quark and the lightest neutralino as possible \lq decay chain' for the produced squarks, it is not necessary to describe all other partial decay widths differentially. Therefore, they can be extracted from the literature or their implementation in \textsc{Sdecay}. In order to implement the various decay routines from \textsc{Sdecay} in our code the following adaptions had to be made for the individual decay modes:

\begin{itemize}
 \item Electroweak decays: $\sq_i \rightarrow q_i \tilde{\chi}^0_k$ ($k=1,2,3,4$) and $\sq_i \rightarrow q_j \tilde{\chi}^\pm_l$ ($l=1,2$)\\
   The decays into neutralinos $\tilde{\chi}^0_k$ or charginos $\tilde{\chi}^\pm_l$ are mediated by electroweak interactions. The decay into charginos is only possible for left-chiral squarks. In the routines for the (N)LO results \cite{decay1} taken from \textsc{Sdecay} only the conventions for the parameters, especially those entering the calculation of the squark-quark-gaugino vertex, had to be adapted.   
   In our calculation the weak mixing angle $\theta_W$ is determined according to Eq.~(10.11) from \cite{pdg}, yielding
\be
\sin^2\theta_W = \frac{1}{2}-\sqrt{\frac{1}{4}-\frac{\pi\alpha(m_Z)}{\sqrt{2} G_F m_Z^2}}\, .
\ee
All other parameters needed for the numerical evaluation of the decay widths can be found in Sec.~\ref{sec:setup}.

 \item Strong decay: $\sq_i \rightarrow q_i \go$\\
 The NLO corrections to this strong decay mode have been calculated in \cite{decay3}. However, this calculation has been performed for degenerate squark masses and implemented in the same way in \textsc{Sdecay}. 
 In order to incorporate the full mass dependence, we have calculated the gluino self energy using the corresponding function from the calculation of the stop decays \cite{decay2}.
 In these decays the correct $\tilde{t}_{1,2}$ masses, the top quark mass and the $\tilde{t}$ mixing angles have been used. For each squark of the first two generations this function is called by replacing the appropriate squark mass and setting the quark mass and mixing angles to zero.
 
 Also in the calculation of $\alpha_s$, where the heavy particles are decoupled from the running, the squark masses are assumed to be degenerate.
  To restore the full mass dependence in $\alpha_s$, the logarithms of the masses of the heavy, decoupled particles have been modified to obtain the logarithms given in Eq.~(3) of \cite{ownpaper}. In \textsc{Sdecay} the strong coupling constant is converted from the $\overline{\textnormal{MS}}$ scheme, used in the original calculation, to the $\overline{\textnormal{DR}}$ scheme. In order to use $\alpha_s$ as implemented in the \PB\ we calculate $\alpha_s$ in the $\overline{\textnormal{MS}}$ scheme by omitting the conversion factor introduced in \textsc{Sdecay}.

\end{itemize}

\subsection{Combination with the Production Processes}
\label{sec:proddec}
A consistent combination of the production processes at NLO with the subsequent decays of the squarks, $\sq \to q + \neutone$, or antisquarks, $\bar{\sq} \to \bar{q} + \neutone$, at NLO is the next necessary step.
In this combination we take into account only those contributions to the process $ p p \to 2 q + 2 \tilde{\chi}_1^0$ that lead to two on-shell intermediate squarks. In the narrow width approximation, which is valid in the scenarios analysed here since the widths of the squarks fulfil $\Gamma_{\tilde{q}_i}/m_{\tilde{q}_i}\ll 1$, the differential cross section factorizes into the production cross section times the branching ratios of both squark decays
\be
 \ds_{\textnormal{tot}} = \ds_{\textnormal{prod}} \ \frac{\dGa}{\Ga} \ \frac{\dGb}{\Gb}\, .
 \label{eq:LOproddec}
\ee
By applying the narrow width approximation we not only neglect
contributions with off-shell squarks, which are known to be suppressed
by $\Gamma_{\tilde{q}_i}/m_{\tilde{q}_i}$, but also  non-factorizable
higher-order contributions. The latter ones comprise interactions between particles of the production and decay stage or between final-state particles of the two decays. These contributions are expected to be suppressed by $\Gamma_{\tilde{q}_i}/m_{\tilde{q}_i}$ as well \cite{khoze1,khoze2}. Only long-range interactions induced by the exchange of soft gluons could still affect the results of exclusive observables. However, an analysis of these effects is beyond the scope of this publication. 

Aiming at a combination of the decays at NLO with the production process at NLO the factors in Eq.~(\ref{eq:LOproddec}) have to be replaced by the NLO quantities:
\be
 \label{eq:NLOquantities}
  \ds_{\textnormal{tot}} = (\ds_0 + \alpha_s \ds_1) \  \frac{\textnormal{d}\Gamma^{\sq_1 \rightarrow \neutone q}_0 + \alpha_s \textnormal{d}\Gamma^{\sq_1 \rightarrow \neutone q}_1}{\Gamma_{\textnormal{tot},0}^{\sq_1} + \alpha_s \Gamma_{\textnormal{tot},1}^{\sq_1} }  \   \frac{\textnormal{d}\Gamma^{\sq_2 \rightarrow \neutone q}_0 + \alpha_s \textnormal{d}\Gamma^{\sq_2 \rightarrow \neutone q}_1}{\Gamma_{\textnormal{tot},0}^{\sq_2} + \alpha_s \Gamma_{\textnormal{tot},1}^{\sq_2} }\ . 
\ee
This expression obviously includes beyond-NLO contributions. In order to strictly consider NLO accuracy it has to be expanded to NLO in $\alpha_s$. There exist two approaches for this problem, both developed in the context of single and pair production of top quarks \cite{proddec2,proddec1}.

 In the first approach a Taylor expansion of the full expression is performed. This leads to a formula which is normalized to the LO total widths and subtracts the ratios of the NLO corrections to the total widths over the LO total widths from the first term:
 \begin{eqnarray}
 \label{eq:proddec1}
 \ds_{\textnormal{tot}} &=& \,\,\,\frac{1}{\GaLO \GbLO} \Biggl[\ds_0\, \dGa_0 \,\dGb_0  \left(1-\frac{\alpha_s \GaNLO }{\GaLO }-\frac{\alpha_s \GbNLO }{\GbLO }\right) \\
 &+ & \alpha_s \left( \ds_0\, \dGa_1 \dGb_0 + \ds_0\, \dGa_0 \dGb_1 + \ds_1\, \dGa_0 \dGb_0  \right)\Biggr]\, .\nonumber
 \end{eqnarray}
This subtracted term might lead to negative contributions, if the NLO corrections to the total width are positive and large while the corrections to the partial widths are small. However, this expansion has the advantage that the sum over all possible decay channels reproduces the production cross section, {\it i.e.} the branching ratios of all subchannels add up to one.

 In the second approach only the numerator is expanded in $\alpha_s$ while the NLO total widths are kept in the denominator. This expansion avoids the problem of potentially negative contributions and leads to the expression:
  \begin{eqnarray}
 \label{eq:proddec2}
 \ds_{\textnormal{tot}} &= &\,\,\,\frac{1}{\Ga \Gb} \Biggl[\ds_0 \,\dGa_0 \dGb_0   + \alpha_s \bigl( \ds_0\, \dGa_1 \dGb_0 \nonumber\\
                        & &+ \ds_0\, \dGa_0 \dGb_1 + \ds_1\, \dGa_0 \dGb_0  \bigr)\Biggr].
  \end{eqnarray}
In this approach summing over all possible decay channels does not reproduce the production cross section, as the branching ratios do not add up to one, and in this sense unitarity is violated.

Since both expansions to NLO accuracy may cause problems the complete expression in Eq.~(\ref{eq:NLOquantities}) can be used as an alternative approach. On the one hand, in this approach the branching ratios sum up to one, but on the other hand only parts of the possible beyond-NLO corrections are included. Given a good convergence of the perturbative series we expect these terms to be small, however.

In chapter \ref{ch:res} results for all three possible combinations of production and decays at NLO, according to Eqs.~(\ref{eq:NLOquantities})-(\ref{eq:proddec2}), will be presented and compared. However, in all other results the Taylor expansion of the cross section, Eq.~(\ref{eq:proddec1}), will be used since in this approach the unitarity of branching ratios is preserved. In the scenarios analysed here the subtracted terms in Eq.~(\ref{eq:proddec1}) are unproblematic, {\it i.e.} the NLO corrections to the total decay widths are small (see Tab.~\ref{tab:totwidths}).
\section{Implementation and Results}
\label{ch:res}
After a brief discussion of the steps required for the implementation of squark production and decay in the \PB~this section summarizes our main findings, including both numerical results at fixed order perturbation theory and after application of different parton showers. Moreover, we present some results for total rates after applying realistic experimental search cuts.
\subsection{Implementation in the \PB}
The implementation of squark-antisquark production in the \PB~is essentially identical to the case of squark pair production, which has been extensively discussed in \cite{ownpaper}. Besides several changes in the main code required for the consideration of processes with strongly interacting SUSY particles the process-dependent parts have to be provided, which comprise
\bi
\item all independent flavour structures contributing to the Born and real channels,
\item the Born and the spin/colour-correlated matrix elements squared,
\item the finite part of the virtual contributions,
\item the real matrix elements squared
\item and the colour flows for the Born configurations.
\ei
The implementation of the various subtraction schemes discussed in Sec.~\ref{sec:virtreal} is rather involved and has been described in detail in \cite{ownpaper}, too. In essence, we have implemented (besides the two DR schemes) several versions of the DS scheme by splitting the real matrix element squared into a part containing the resonant gluino contributions and the corresponding subtraction terms and a part containing all other terms. The resonant parts do not contain any IR singularities and can therefore be treated independently from the \textsc{Powheg}-like event generation, similar to the \lq hard' part $\matR_h$ of the real matrix elements squared introduced below.

We have implemented these building blocks for squark-antisquark production into the version~2 (V2) of the \PB~and ported our previous implementation of squark pair production to the V2.
This newer version of the \PB~allows for the consideration of NLO corrections to the decays of the on-shell produced squarks. We use this new option to combine our results for the NLO production processes with the corrections to the specific decay $\sq\rightarrow q \neutone$ described in the previous section. Besides taking into account the decay products in the flavour structures as described in the manual of the \PB~V2 this requires the combination of the production and decay matrix elements according to the combination formula in Eq.~(\ref{eq:proddec1}). Moreover, the FKS subtraction of the IR divergences related to the gluon emission off either the squark or the quark in the NLO corrections to the decay process requires the specification of the colour correlated Born matrix elements squared. These are trivial in the case at hand and read in the convention of the \PB~$ \matB_{\sq q} = C_F \matB$.

 In order to check the correctness of the implemented results the same tests as described in \cite{ownpaper} have been performed. These comprise a comparison of numerous differential distributions evaluated at NLO with the corresponding results after generation of the hardest emission according to the \textsc{Powheg} method, both at the level of the production processes and after including the decays. While we find an excellent agreement for inclusive quantities, a strong enhancement of the \textsc{Powheg} results compared to the respective NLO distributions is observed for exclusive quantities like the transverse momentum of the squark-antisquark system, $p_T^{\sq\sqbar}$. The same artificial enhancement has already been observed in case of squark pair production and can be cured by using the soft/collinear limits of the real matrix elements squared $\matR$ instead of the full expressions for the generation of the hardest radiation. In the \PB~this is achieved by introducing a function $\matF$ which separates the soft/collinear part $\matR_s$ and the hard part $\matR_h$ of the real matrix elements squared:
 
 \be
 \matR = \matF \matR + (1-\matF) \matR \equiv \matR_s + \matR_h\, .
\ee
This function $\matF$ has to fulfil $\matF\rightarrow 1$ in the soft/collinear limit and should vanish far away from the corresponding phase space regions. In the \PB~the functional form 
\be
  \matF=\frac{h^2}{p_T^2+h^2}
  \label{eq:fdamp}
\ee
is used, with the transverse momentum $p_T$ of the emitted parton with respect to the emitter and a damping parameter $h$ (see \cite{nason} and \cite{powhegbox} for further details). Similar to our earlier studies on squark pair production we use $h=50\,\text{GeV}$ throughout. This choice was found to damp the artificial enhancement in the $p_T^{\sq\sqbar}$ distribution and reproduces the NLO prediction for $p_T^{\sq\sqbar}\gtrsim 200\,\text{GeV}$, while maintaining the Sudakov damping for small transverse momenta inherent in the \textsc{Powheg} method. 
  

\subsection{Setup}
\label{sec:setup}
For the numerical analysis we consider two mSUGRA scenarios which are not yet excluded by data, see {\it e.g.} \cite{atlasexcl4,cmsexcl6}. The scenarios are based on the CMSSM points $10.3.6^*$\footnote{For the point $10.3.6$ $m_0$ has been modified to get a mass spectrum consistent with the latest exclusion bounds.} and $10.4.5$ from \cite{susybench}. The input parameters of these scenarios are summarized in Tab.~\ref{tab:msugra}. The mass spectrum of the SUSY particles has been generated with \textsc{Softsusy 3.3.4} \cite{softsusy}, the resulting on-shell masses are then used as input parameters. For the SM parameters the following values are used \cite{pdg}:
\begin{gather}
  m_Z = 91.1876\, \text{GeV}, \quad G_F=1.16637\cdot 10^{-5}\, \text{GeV}^{-2},  \nonumber\\
  \alpha_{em}(m_Z)=1/127.934,\quad  \alpha_s(m_Z)=0.118,\\
  m_b^{\overline{\textnormal{MS}}}(m_b)=4.25\, \text{GeV}, \quad m_t=174.3\, \text{GeV}, \quad m_{\tau} = 1.777\, \text{GeV} , \quad m_{c}^{\overline{\textnormal{MS}}}(m_c) = 1.27\, \text{GeV}.\nonumber
\end{gather}
  
\begin{table}
\bc
\begin{tabular}{|c |c | c | c | c | c|}
  \hline
  Scenario & $m_0$  &  $m_{1/2}$  & $A_0$   &  $\tan(\beta)$  & $\text{sgn}(\mu)$\\\hline\hline
  $10.3.6^*$  & $825 \, \text{GeV}$  &  $550\, \text{GeV}$  & $0\, \text{GeV}$ &    $10$        &    $+1$       \\
  $10.4.5$  & $1150\, \text{GeV}$  &  $690\, \text{GeV}$  & $0\, \text{GeV}$ &    $10$        &    $+1$       \\\hline
\end{tabular}
  \caption{\label{tab:msugra}The input parameters for the considered scenarios.}
\ec
\vspace*{-0.2cm}
\end{table}

As \textsc{Softsusy} implements non-vanishing Yukawa corrections, there is a small difference between the masses of the second-generation squarks and the corresponding first-generation ones, {\it i.e.} $m_{\sul}\neq m_{\scl}$ etc. To simplify the analysis and save computing time these masses are replaced by the mean of the mass pairs, {\it i.e.} $m_{\sul}$ and $m_{\scl}$ are replaced by $(m_{\sul}+m_{\scl})/2$ and so on. The obtained masses for the squarks of the first two generations and the gluino masses are summarized in Tab.~\ref{tab:sqmass}. 
Note that for the point $10.3.6^*$ the mass hierarchy is $m_{\sq}>m_{\go}$, while for $10.4.5$ $m_{\sq}<m_{\go}$, the latter point requiring the subtraction of contributions with on-shell intermediate gluinos as described in Sec.~\ref{sec:virtreal}. 
Here, the DS$^*$ method is used, with a default value for the regulator $\Gamma_{\go}=1\,\text{GeV}$ (recall that this regulator is only needed if a subtraction is required, thus in all other cases it is set to zero).

\begin{table}[!ht]
\renewcommand{\arraystretch}{1.3}
\bc
\begin{tabular}{| c |c | c | c | c | c| }
  \hline
  Scenario & $m_{\tilde{u}_L} = m_{\scl}$ & $m_{\tilde{u}_R} = m_{\tilde{c}_R}$ & $m_{\tilde{d}_L} = m_{\ssl}$ & $m_{\tilde{d}_R} = m_{\ssr}$ & $m_{\tilde{g}}$ \\\hline\hline

  $10.3.6^*$  & $1799.53$ & $1760.21$  & $1801.08$ & $1756.40$ & $1602.96$ \\

  $10.4.5$  & $1746.64$ & $1684.31$  & $1748.25$ & $1677.82$ & $1840.58$ \\\hline
\end{tabular}
\caption{\label{tab:sqmass}The squark masses in $\text{GeV}$ obtained with the parameters from Tab.~\ref{tab:msugra} after averaging the masses of the first two generations as described in the text.}
\ec
\vspace*{-0.2cm}
\end{table}

%
%
Furthermore, the partial and total decay widths of the squarks depend on the masses of the charginos and neutralinos and the respective mixing matrices.
The masses of the neutralinos and charginos for the two scenarios are given in Tab.~\ref{tab:neutmass}.
\begin{table}
\renewcommand{\arraystretch}{1.2}
\bc
\begin{tabular}{| c |c | c | c | c | c| c| }
  \hline
  Scenario & $m_{\neutone}\, [\text{GeV}]$ & $m_{\neuttwo}\, [\text{GeV}]$ & $m_{\neutthree}\, [\text{GeV}]$ &  $m_{\neutfour}\, [\text{GeV}]$ & $m_{\charone}\, [\text{GeV}]$ & $m_{\chartwo}\, [\text{GeV}]$ \\\hline\hline

  $10.3.6^*$  & $290.83$ & $551.76$  & $-844.74$ & $856.87$ & $551.99$ & $856.40$ \\

  $10.4.5$  & $347.71$ & $657.84$  & $-993.42$ & $1003.79$ & $856.06$ & $1003.46$\\\hline
\end{tabular}
\caption{\label{tab:neutmass}The neutralino and chargino masses for the benchmark scenarios defined in Tab.~\ref{tab:msugra}.}
\ec
\vspace*{-0.2cm}
\end{table}
The neutralino mixing matrices for the scenarios $10.3.6^*$ and $10.4.5$ read
\begin{equation}
\begin{aligned}
N^{10.3.6^*}&=
\left( \begin{array}{rrrr}
 0.99759 & -0.00979 & 0.06292 & -0.02740 \\
 0.02329 &  0.97889 &-0.16595 &  0.11704 \\
-0.24682 & 0.03551 & 0.70512 &  0.70776 \\
-0.06044 & 0.20106 & 0.68651 & -0.69615 \end{array} \right)\qquad\text{and}\\\\
N^{10.4.5} &= 
\left( \begin{array}{rrrr}
 0.98267 & -0.00716 & 0.05338 & -0.02358 \\
-0.20847 &  0.02997 & 0.70567 &  0.70760 \\
 0.01724 &  0.98393 &-0.14473 &  0.10318 \\
-0.05226 &  0.17590 & 0.69154 & -0.69865 \end{array} \right).
\end{aligned}
\end{equation}
In order to diagonalize the chargino mass matrix two matrices are needed, one for the left-handed components (denoted $U$) and one for the right-handed ones (denoted $V$).
These $2\times2$ mixing matrices are parametrized as ($i=U,V$)
\be
\left(
\begin{array}{cc}
 \cos{\theta_i} & -\sin{\theta_i}\\
 \sin{\theta_i} & \cos{\theta_i}
\end{array}
\right)\, .
\ee
The mixing angles are given by $\cos{\theta_U} = 0.97213$ and $\cos{\theta_V} = 0.98594$ for the parameter point $10.3.6^*$. Likewise, those for the scenario $10.4.5$ read $\cos{\theta_U} = 0.97894$ and $\cos{\theta_V} = 0.98914$.

The renormalization ($\mu_R$) and factorization ($\mu_F$) scales are chosen as $\mu_R=\mu_F=\overline{m}_{\sq}$, with $\overline{m}_{\sq}$ representing the average of the squark masses of the first two generations. For the two scenarios defined above one obtains $\overline{m}_{\sq}^{10.3.6^*}=1779.31\,\text{GeV}$ and $\overline{m}_{\sq}^{10.4.5}=1714.25\,\text{GeV}$, respectively. All subchannels for the production of first- and second-generation squarks are taken into account for the results, {\it i.e.}\ if not stated otherwise all results presented in the rest of this section are obtained by adding up the subchannels. For squark pair production the (tiny) contributions of the antisquark pair production channels are always taken into account.

The PDFs are taken from the \textsc{LHAPDF} package \cite{lhapdf}. For the LO results shown in the following the LO set \textsc{CTEQ6L1} \cite{cteq6} with $\alpha_s(m_Z)=0.130$ is used, while the NLO results are calculated with the NLO set \textsc{CT10NLO} with $\alpha_s(m_Z)=0.118$ \cite{cteq}. The strong coupling constant for the LO results is correspondingly computed using the one-loop renormalization group equations (RGEs), while the value used in the NLO results is obtained from the two-loop equations.

All results are calculated for the LHC with $\sqrt{s} = 14\,\text{TeV}$. The error bars shown in the following represent the statistical errors of the Monte Carlo integration.

Taking into account the decays of the produced squarks into $q\neutone$ or applying a parton shower algorithm leads to a potentially large number of partons in the final state. These partons are clustered into jets with \textsc{Fastjet 3.0.3} \cite{fastjet1,fastjet2}. To this end the anti-$k_T$ algorithm \cite{antikt} is adopted, using $R=0.4$.
In the following only minimal cuts are applied on the transverse momentum and the pseudorapidity of the resulting jets:
\be
   p_T^j>20\,\text{GeV}\quad \textnormal{and} \quad |\eta^j|<2.8\, .
\ee
Except for the results shown in Sec.~\ref{sec:totrat} no event selection cuts are imposed.

\subsection{Numerical Results}
\subsubsection{Results at Fixed Order}
\label{sec:fixedorder}
The first part of this section is devoted to a discussion of the NLO corrections to squark-antisquark production. In the second part we present some results for the combination of production and decay, both for squark-antisquark and squark pair production. Hence this part extends our previous results for the squark pair production processes in \cite{ownpaper} by also including the NLO corrections to the decay.

\subsubsection*{Squark-Antisquark Production}
The results for the total squark-antisquark production cross sections determined at LO and NLO for the two benchmark scenarios defined in Sec.~\ref{sec:setup} are summarized in Tab.~\ref{tab:totprod}. In order to assess the theoretical uncertainties we vary the renormalization and factorization scales by a factor two around the central value $\mu=\overline{m}_{\sq}$. The resulting percental uncertainties are also given in the table. Considering the resulting $K$-factors we note that in both cases the SUSY-QCD NLO corrections are positive and large, resulting in $K\equiv \sigma_{\rm NLO}/\sigma_{\rm LO} \approx 1.4$. The scale uncertainties are strongly reduced by taking into account the NLO corrections, as expected.
 \begin{table}[t]
\renewcommand{\arraystretch}{1.2}
\bc
\begin{tabular}{|c || c |c  | c |}\hline
Scenario  &  $\sigma_{\text{LO}}^{\pm\Delta \sigma} [\text{fb}]$  &  $\sigma_{\text{NLO}}^{\pm\Delta \sigma} [\text{fb}]$  &  K-factor \\\hline
$10.3.6^*$ & $2.319^{+34\%}_{-24\%} $ & $3.218^{+13\%}_{-14\%}$ & 1.39 \\
$10.4.5$   & $3.098^{+34\%}_{-24\%}$  & $4.366^{+14\%}_{-14\%}$ & 1.41\\
\hline
\end{tabular}
  \caption{\label{tab:totprod}The LO and NLO cross sections for squark-antisquark production for the two benchmark scenarios defined in Sec.~\ref{sec:setup}. The theoretical error estimates $\pm\Delta\sigma$ have been obtained by varying the renormalization and factorization scales by a factor two around the central values.}
\ec
\vspace*{-0.2cm}
\end{table}
Turning next to the individual $K$-factors for the subchannels contributing to squark-antisquark production we observe that they differ significantly from the total $K$-factor obtained after summing the cross sections for all individual channels. To illustrate this point, the LO/NLO cross sections and the resulting $K$-factors for the production channels involving only squarks of the first generation are given in Tab.~\ref{tab:totxs} for the CMSSM point $10.3.6^*$. Note, that the channels with squarks of the same flavour and chirality in the final state, displayed in the first four rows of the table, have contributions from $gg$ initial states and therefore larger $K$-factors than channels with squarks of different flavour or chirality. Hence the assumption that the individual $K$-factors can be approximated by the total $K$-factor obtained from \textsc{Prospino} is in general not valid. 
 \begin{table}[t]
\renewcommand{\arraystretch}{1.2}
\bc
\begin{tabular}{|c || c |c  | c |}\hline
Process  &  $\sigma_{\text{LO}} [\text{fb}]$  &  $\sigma_{\text{NLO}} [\text{fb}]$  &  K-factor \\\hline\hline
$\tilde{u}_{L}\bar{\tilde{u}}_{L}$  & $  9.51\cdot 10^{-2} $ & $ 1.43\cdot 10^{-1} $ & $ 1.50 $ \\ 
$\tilde{u}_{R}\bar{\tilde{u}}_{R}$  & $  1.14\cdot 10^{-1} $ & $ 1.72\cdot 10^{-1} $ & $ 1.51 $ \\ 
$\tilde{d}_{L}\bar{\tilde{d}}_{L}$  & $  5.50\cdot 10^{-2} $ & $ 8.79\cdot 10^{-2} $ & $ 1.60 $ \\ 
$\tilde{d}_{R}\bar{\tilde{d}}_{R}$  & $  6.89\cdot 10^{-2} $ & $ 1.11\cdot 10^{-1} $ & $ 1.61 $ \\ 
$\tilde{u}_{L}\bar{\tilde{u}}_{R}$  & $  3.75\cdot 10^{-1} $ & $ 5.12\cdot 10^{-1} $ & $ 1.37 $ \\ 
$\tilde{d}_{L}\bar{\tilde{d}}_{R}$  & $  1.41\cdot 10^{-1} $ & $ 1.70\cdot 10^{-1} $ & $ 1.21 $ \\ 
$\tilde{u}_{L}\bar{\tilde{d}}_{L}$  & $  6.98\cdot 10^{-2} $ & $ 7.89\cdot 10^{-2} $ & $ 1.13 $ \\ 
$\tilde{u}_{L}\bar{\tilde{d}}_{R}$  & $  2.98\cdot 10^{-1} $ & $ 3.54\cdot 10^{-1} $ & $ 1.19 $ \\ 
$\tilde{u}_{R}\bar{\tilde{d}}_{L}$  & $  2.94\cdot 10^{-1} $ & $ 3.49\cdot 10^{-1} $ & $ 1.19 $ \\ 
$\tilde{u}_{R}\bar{\tilde{d}}_{R}$  & $  8.36\cdot 10^{-2} $ & $ 9.54\cdot 10^{-2} $ & $ 1.14 $ \\ 
\hline\hline
Sum  & 1.59 & 2.07 & 1.30   \\\hline
\end{tabular}
  \caption{\label{tab:totxs}The LO and NLO cross sections for squark-antisquark production of the first generation obtained for the CMSSM point $10.3.6^*$. The charge conjugate channels have been combined.}
\ec
\vspace*{-0.2cm}
\end{table}
Determining the individual corrections consistently is especially important if the decays are taken into account and the branching ratios of the different squarks differ significantly for the specific decay channel under consideration. In order to assess the possible numerical impact of this approximation we consider the decay $\sq\rightarrow q \neutone$ at LO at the level of total cross sections, {\it i.e.} we multiply the production cross sections for the individual squark-antisquark production channels with the respective LO branching ratios. In this step we take into account the contributions of the second generation squarks as well.

We first consider the benchmark scenario 10.3.6$^*$. Using the correctly calculated NLO results for the individual production channels, multiplying them with the corresponding branching ratios and summing all channels, we obtain
\be
\sum_{\text{channels}}\sigma_{\text{NLO}}\cdot \text{BR}^{\text{LO}}\left(\sq\rightarrow \neutone q \right) \cdot \text{BR}^{\text{LO}}\left(\sqbar\rightarrow \neutone \qbar \right)
  =0.139\, \text{fb}.
\ee

To mimic the way \textsc{Prospino} obtains the individual NLO results a common $K$-factor has to be calculated, using an averaged squark mass $m_{\sq} = 1779.31 \,\text{GeV}$. In the case at hand this leads to
\begin{equation}
\begin{aligned}
  \sigma^{\text{avg}}_{\text{LO}} = 2.315 &\,\text{fb}\; ,\qquad \sigma^{\text{avg}}_{\text{NLO}} = 3.218\,\text{fb} \\
  &\Rightarrow  K^{\text{avg}} =1.39\,.
\end{aligned}  
\end{equation}
Note that the difference compared to the full calculation given in Tab.~\ref{tab:totprod} is marginal and not visible when rounding to the second decimal place. This is due to the fact that the spread in the squark masses is rather small. 
Multiplying the LO result for each subchannel with this common $K$-factor and the corresponding branching ratios gives
\be
\sum_{\text{channels}}\sigma_{\text{LO}}\cdot K^{\text{avg}}\cdot \text{BR}^{\text{LO}}\left(\sq\rightarrow \neutone q \right) \cdot \text{BR}^{\text{LO}}\left(\sqbar\rightarrow \neutone \qbar \right)
  =0.126\, \text{fb}\, .
\ee
Thus the rate obtained with the approximation relying on a constant $K$-factor for all subchannels is roughly $10\%$ smaller for this special case. 

Repeating this procedure for the benchmark scenario $10.4.5$ one obtains for the \textsc{Prospino}-like $K$-factor 
\begin{equation}
\begin{aligned}
    \sigma^{\text{avg}}_{\text{LO}} = 3.090&\,\text{fb}\; , \qquad \sigma^{\text{avg}}_{\text{NLO}} = 4.356\,\text{fb} \\
  &\Rightarrow  K^{\text{avg}}=1.41\,.
\end{aligned}  
\end{equation}
Again, comparing this result to the full calculation given in Tab.~\ref{tab:totprod} the discrepancy is only marginal.

Considering the individual subchannels with the correct individual NLO corrections yields
\be
\sum_{\text{channels}}\sigma_{\text{NLO}}\cdot \text{BR}^{\text{LO}}\left(\sq\rightarrow \neutone q \right) \cdot \text{BR}^{\text{LO}}\left(\sqbar\rightarrow \neutone \qbar \right)
  =0.916\, \text{fb},
\ee
while the approximation of the common $K$-factor gives
\be
\sum_{\text{channels}}\sigma_{\text{LO}}\cdot K^{\text{avg}}\cdot \text{BR}^{\text{LO}}\left(\sq\rightarrow \neutone q \right) \cdot \text{BR}^{\text{LO}}\left(\sqbar\rightarrow \neutone \qbar \right)
  =0.807\, \text{fb}
\ee
and thus again a discrepancy of about $10\%$. 

\subsubsection*{Squark Production and Decay at NLO}
As discussed in Sec.~\ref{sec:proddec} we have used three different approaches to combine the production and decay processes at NLO, differing in the way the combined expression is expanded in $\as$. All approaches require the calculation of the total squark width, either at LO or NLO accuracy. The results for the two considered benchmark scenarios are summarized in Tab.~\ref{tab:totwidths}. 

\begin{table}[ht]
\renewcommand{\arraystretch}{1.2}
\bc
\begin{tabular}{|c || c |c  || c | c | }\hline
  & \multicolumn{2}{c||}{$10.3.6^*$} & \multicolumn{2}{c|}{$10.4.5$} \\\cline{2-5}
  & $\Gamma_{\text{LO}} [\text{GeV}]$ & $\Gamma_{\text{NLO}} [\text{GeV}]$ & $\Gamma_{\text{LO}} [\text{GeV}]$ & $\Gamma_{\text{NLO}}[\text{GeV}] $\\\hline
$\sul$   &  22.79 &  23.44  & 16.21 & 15.81\\ 

$\sur$   &  6.561 &  7.413  &  3.493 & 3.411\\ 

$\sdl$   &  22.78 &  23.45  & 16.14  &  15.74\\ 

$\sdr$   & 3.610 &  4.553  & 0.869 & 0.849 \\ \hline

\end{tabular}
  \caption{\label{tab:totwidths} The total widths for first-generation squarks at LO and NLO for the two scenarios considered here. The widths for the second-generation squarks are identical. For the parameters see the main text. The scale for $\as$ has been set to $\mu_R=\overline{m}_{\sq}$.}
\ec
\vspace*{-0.2cm}
\end{table}
In a first step we compare the numerical results obtained with these approaches, both for differential distributions and total cross sections. 
In Fig.~\ref{fig:compapp} the distributions for the transverse momenta of the hardest and the second hardest jet, $p_T^{j_1/j_2}$, their invariant mass  $m^{j_1j_2}$ and the missing transverse energy $\slashed{E}_T$ are depicted for squark pair production using the benchmark scenario $10.3.6^*$. 
Here, App.~1 corresponds to the Taylor expansion according to Eq.~(\ref{eq:proddec1}), whereas in App.~2 only the numerator in the combination formula is expanded, see Eq.~(\ref{eq:proddec2}). Approach 3 is the result obtained without any expansion, {\it i.e.} these distributions include contributions which are formally of beyond-NLO.

The discrepancies between the approaches 1 and 2 can amount to up to $\matO(15\%)$ for the jet distributions and are largest close to threshold, while the results for $\slashed{E}_T$ reflect only the overall discrepancy in the total cross sections, which amounts for this scenario to approximately 4\%.
The distributions obtained with the third approach do not show any large deviations from the results obtained with the other two approaches, but suffer from large statistical fluctuations. These result from the more complicated structure of the phase space integrations.

The total cross sections for the combined production and decay processes as obtained with the approaches 1 and 2 are summarized in Tab.~\ref{tab:proddec}, both for the scenario 10.3.6$^*$ and 10.4.5. Note that the predictions for the LO cross sections are identical in both approaches and have been calculated according to Eq.~(\ref{eq:LOproddec}) using the LO quantities. Comparing the results for the different predictions at NLO reveals only rather small discrepancies $<4\%$ for the total rates for the scenarios considered here. In the rest of this chapter we use exclusively the first approach to combine production and decay processes.

\bfig[t]
  \begin{minipage}{0.48\textwidth}
   \includegraphics[width=\textwidth,height=5cm]{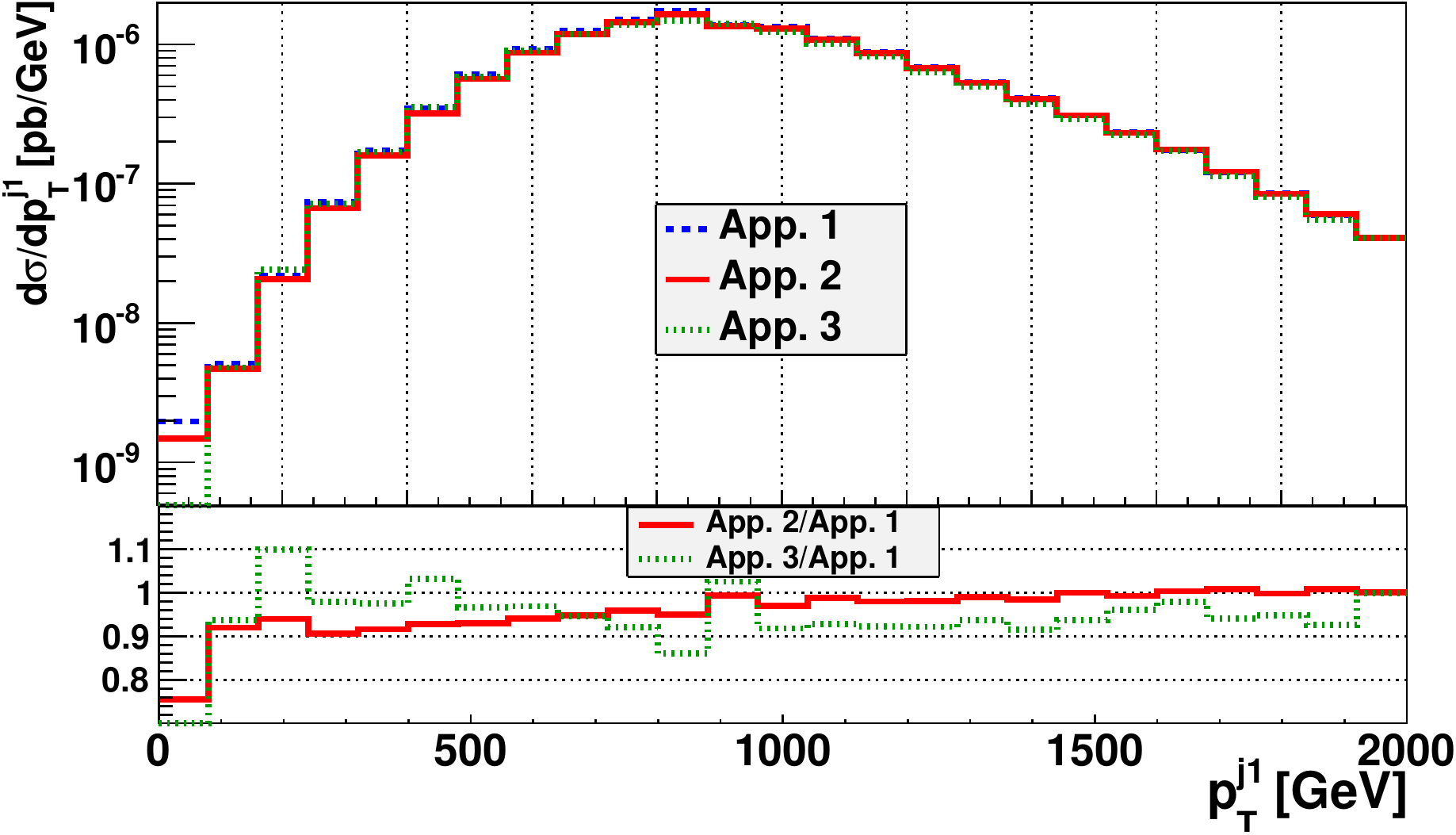}
 \vspace{0.1cm}
 \newline
 \includegraphics[width=\textwidth,height=5cm]{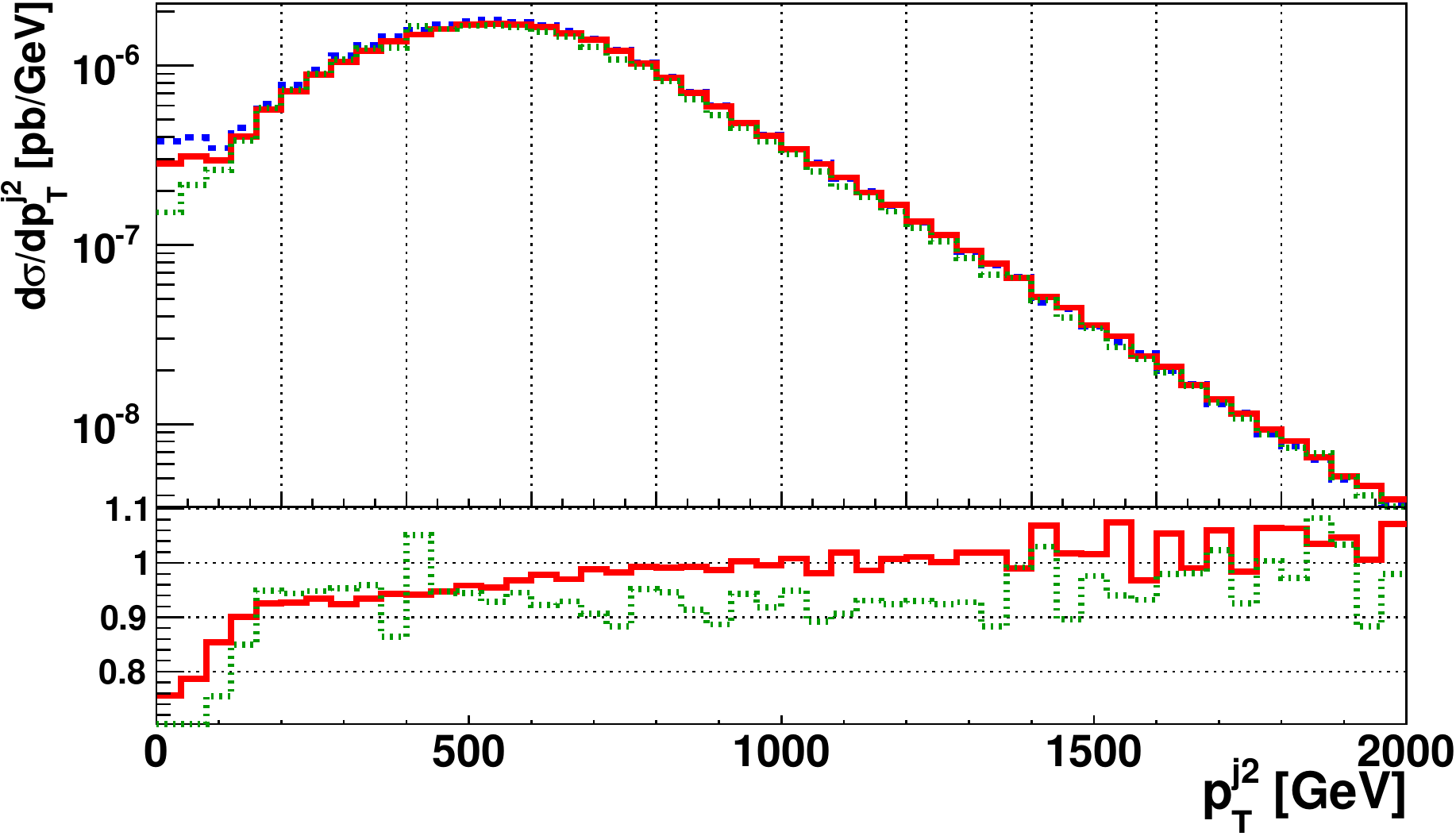}
\end{minipage}
\begin{minipage}{0.04\textwidth}
\end{minipage}
\begin{minipage}{0.48\textwidth}
 \includegraphics[width=\textwidth,height=5cm]{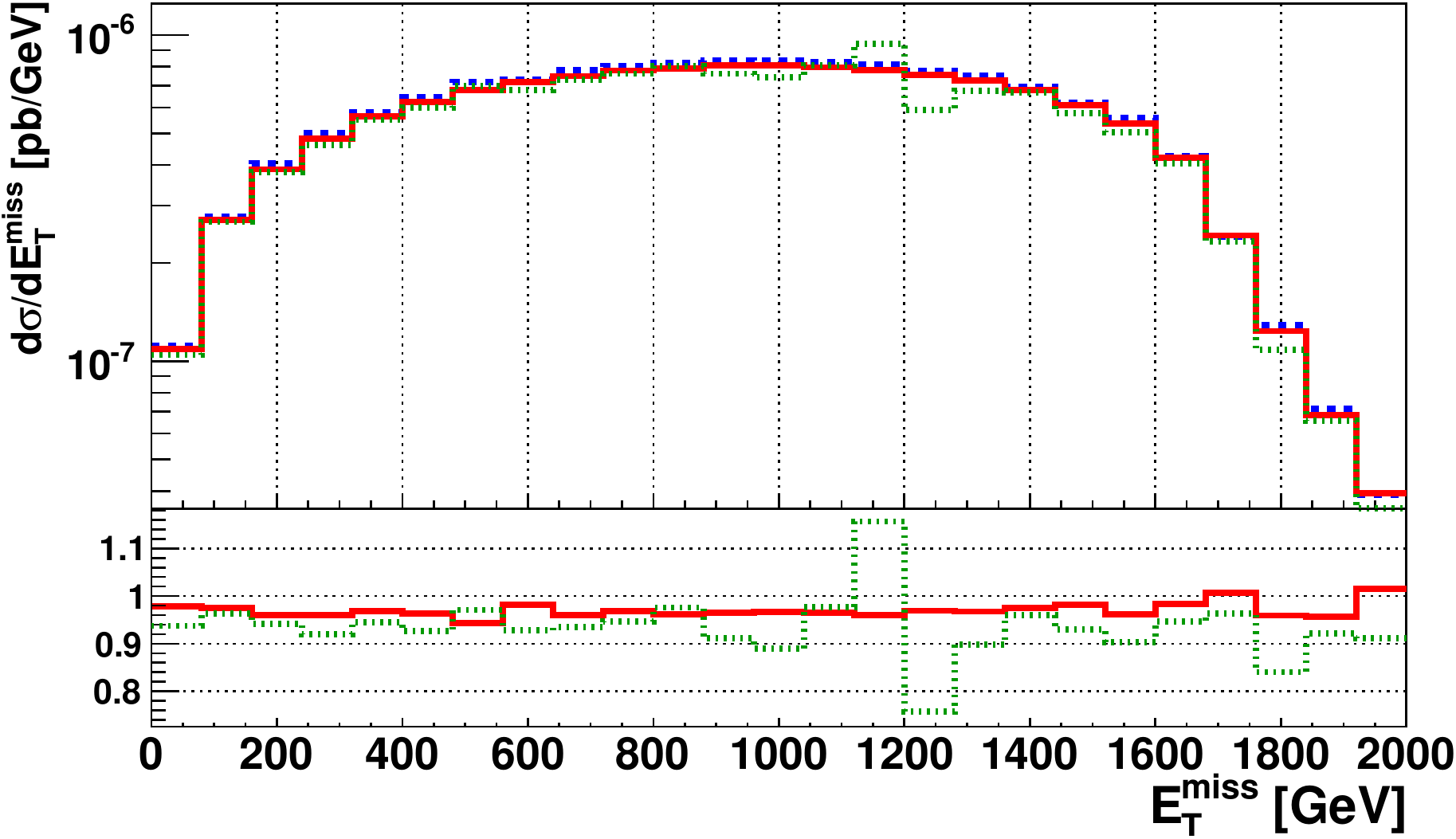}
 \vspace{0.1cm}
 \newline
 \includegraphics[width=\textwidth,height=5cm]{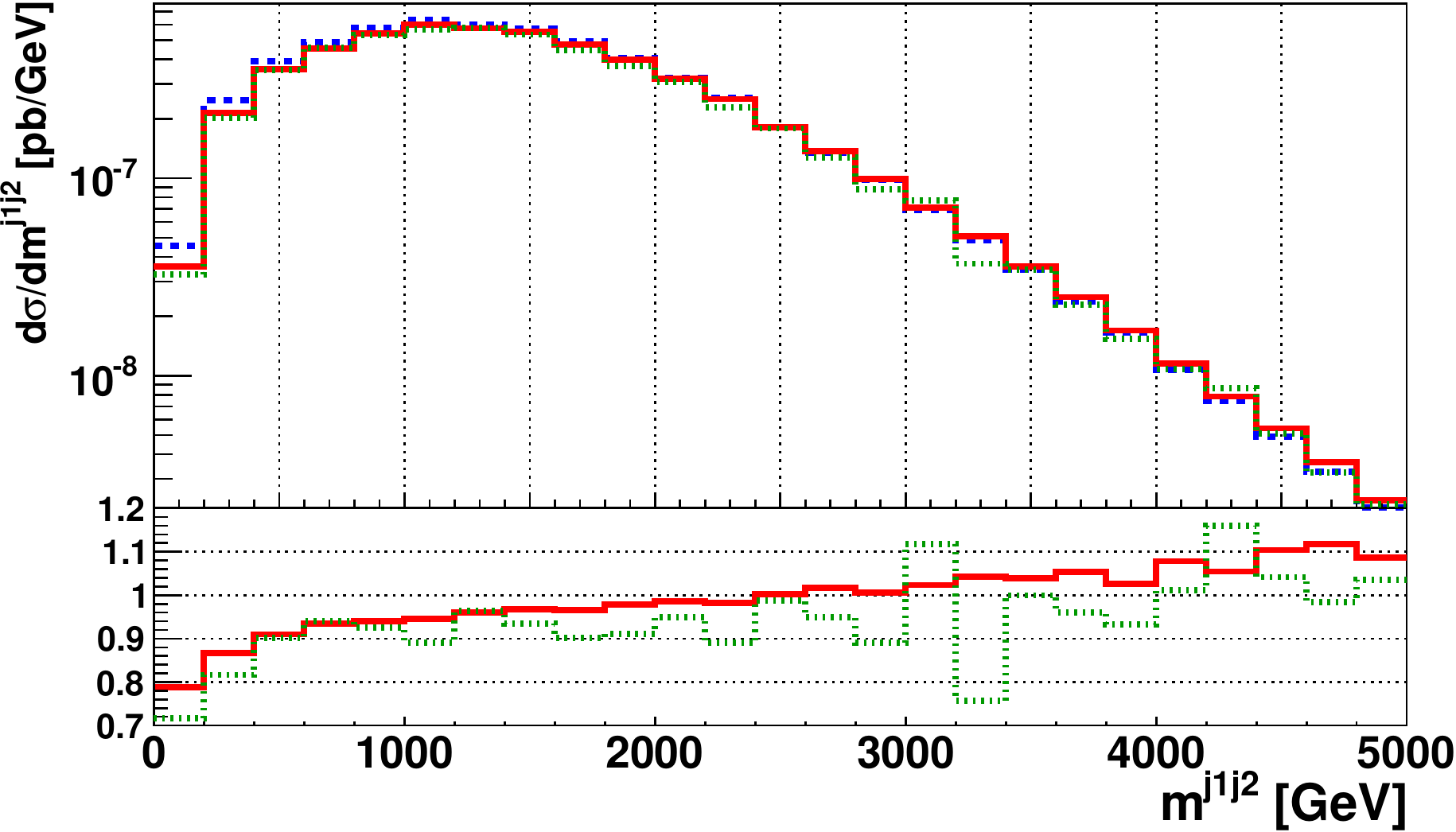}
\end{minipage}
\caption {\label{fig:compapp} Comparison of the three approaches specified in the text for the combination of NLO corrections in production and decay. Shown are the distributions obtained for squark pair production and subsequent decays for the scenario $10.3.6^*$. The lower panels show the differential ratios of the second/third approach with respect to the first approach.}
\efig

\begin{table}
\renewcommand{\arraystretch}{1.2}
\bc
\begin{tabular}{|c || c |c |c  || c | c | c | }\hline
Scenario  & \multicolumn{3}{c||}{$10.3.6^*$} & \multicolumn{3}{c|}{$10.4.5$} \\\hline
Process & $\sigma_{\text{LO}} [\text{fb}]$ & $\sigma_{\text{NLO}} [\text{fb}]$ & $K$-factor &$\sigma_{\text{LO}} [\text{fb}]$ & $\sigma_{\text{NLO}} [\text{fb}]$ & $K$-factor  \\\hline\hline
$\sq\sq$ - App.~1      & 1.34 & 1.12 & 0.84 & 7.57 & 8.75  & 1.16\\
$\sq\sq$ - App.~2      & 1.34 & 1.09 & 0.81 & 7.57 & 8.89  & 1.17\\\hline\hline
$\sq\sqbar$ - App.~1       & $9.29\cdot 10^{-2}$ & $1.03\cdot 10^{-1}$ & $1.11$ & $5.73\cdot 10^{-1}$ & $9.15\cdot 10^{-1}$   & $1.60$\\
$\sq\sqbar$ - App.~2    & $9.29\cdot 10^{-2}$ & $9.88\cdot 10^{-2}$ & $1.06$ & $5.73\cdot 10^{-1}$ & $9.32\cdot 10^{-1}$   & $1.63$\\\hline

\end{tabular}
  \caption{\label{tab:proddec}  Cross sections for squark production and decay at LO and NLO, combined according to Eq.~(\ref{eq:proddec1}) (App.~1) and Eq.~(\ref{eq:proddec2}) (App.~2). }
\ec
\vspace*{-0.2cm}
\end{table}
%
In order to assess the influence of the NLO corrections on differential cross sections we consider in the following the differential $K$-factors for several observables. In Fig.~\ref{fig:LONLOdec1} the LO and the NLO distributions for the transverse momentum of the hardest jet, $p_T^{j_1}$, its rapidity, $y^{j_1}$, the missing transverse energy $\slashed{E}_T$ and the effective mass $m_{\text{eff}} \equiv p_T^{j_1} + p_T^{j_2} + \slashed{E}_T$ are depicted for squark-antisquark production, using the benchmark scenario $10.3.6^*$. The results for the scenario 10.4.5 are qualitatively the same.
Considering the $p_T$ distribution of the hardest jet one observes a strong enhancement of the NLO corrections for small values of $p_T$, while they turn even negative for large values. The result for the second hardest jet, which is not shown here, is qualitatively the same. A similar observation holds for the effective mass: the NLO curve is dragged to smaller values of $m_{\text{eff}}$ and the differential $K$-factor depicted in the lower panel is far from being flat over the whole region. For the $\slashed{E}_T$ predictions, in contrast, the deviation of the differential $K$-factor from the total one is rather small, of $\matO(5\%)$, except for events with very small or very large missing transverse energy. Likewise, the shape of the rapidity distribution of the hardest jet is hardly affected by the NLO corrections. 
\bfig[t]
  \begin{minipage}{0.48\textwidth}
 \includegraphics[width=\textwidth,height=5cm]{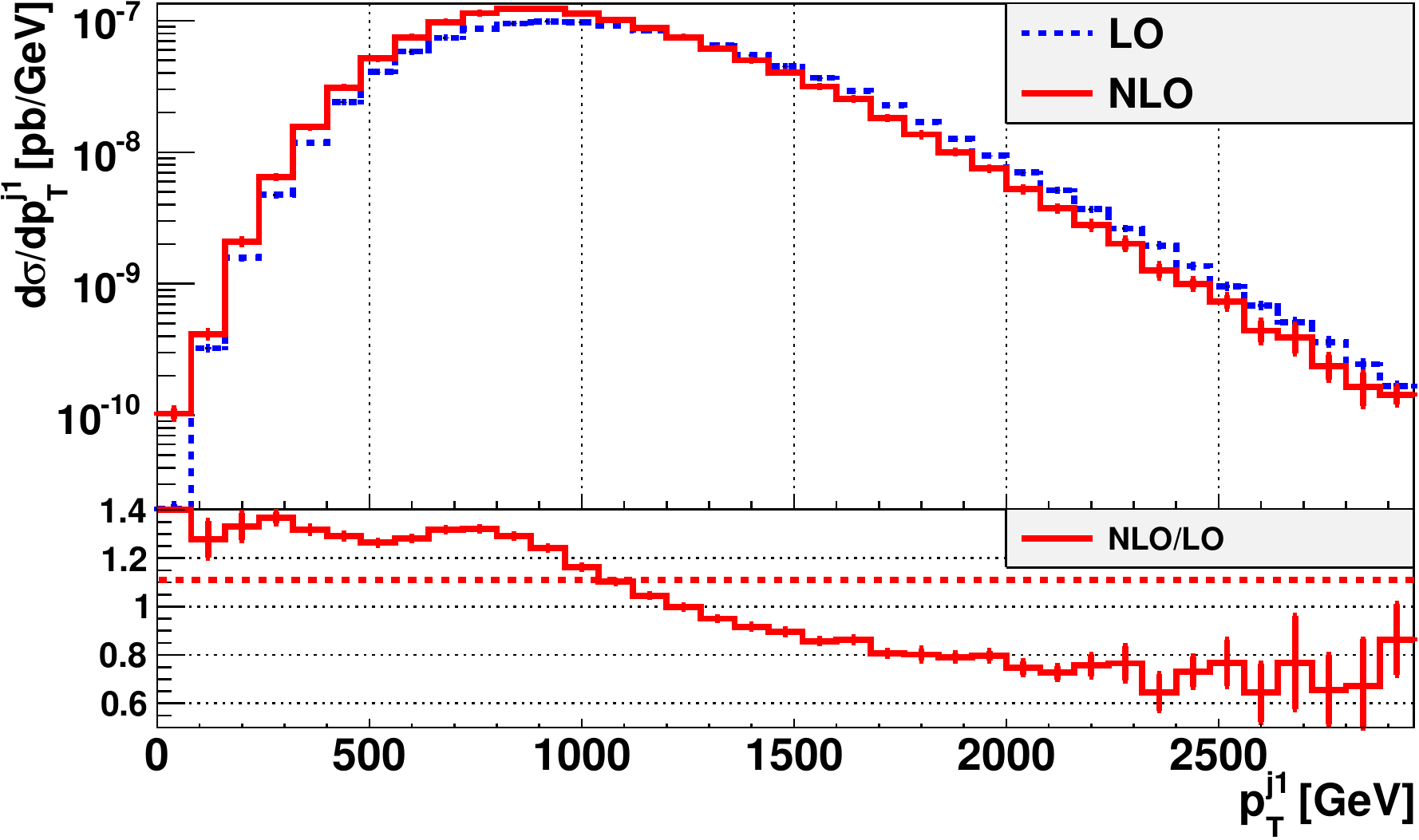}
 \vspace{0.1cm}
 \newline
 \includegraphics[width=\textwidth,height=5cm]{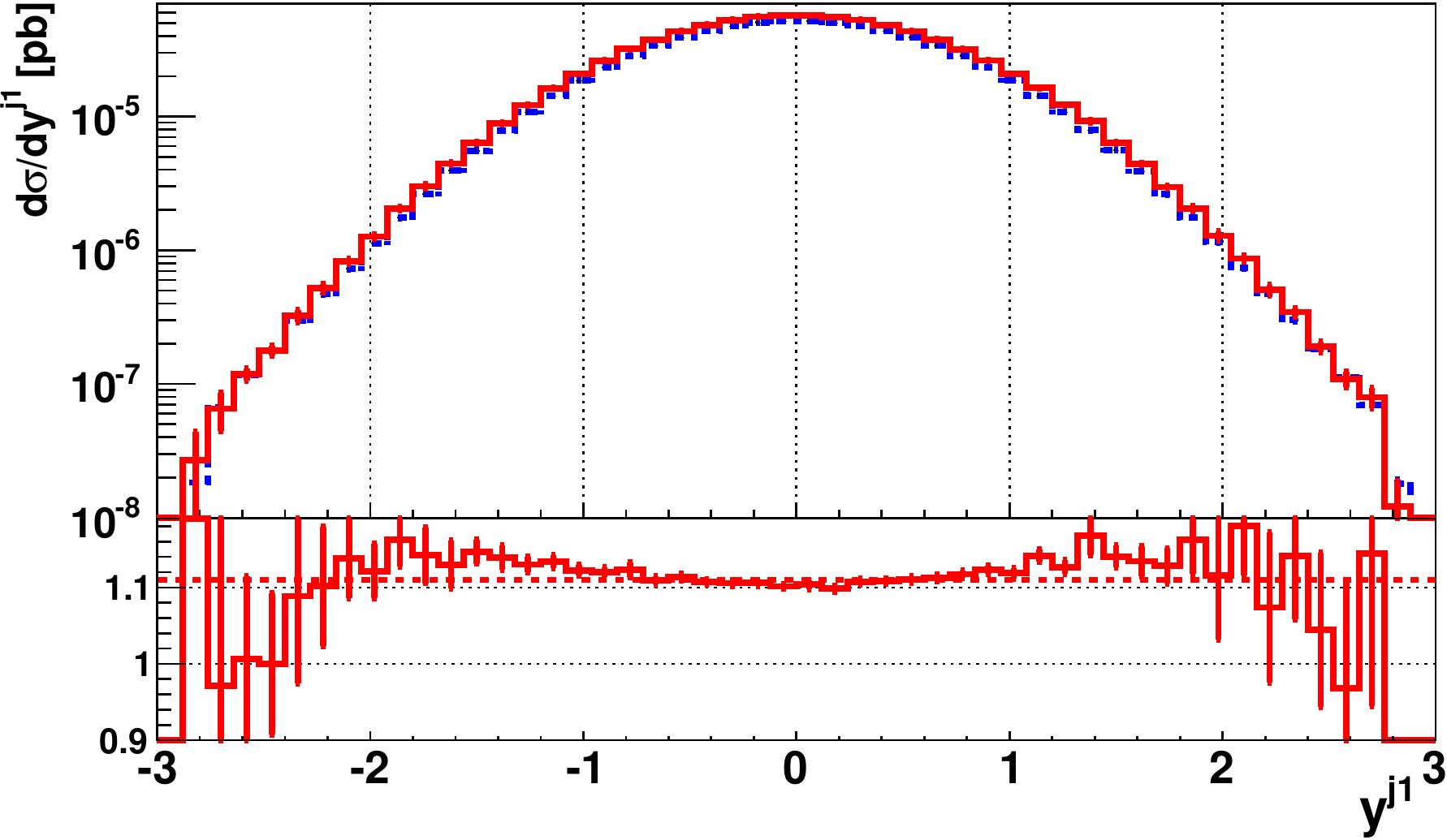}
\end{minipage}
\begin{minipage}{0.04\textwidth}
\end{minipage}
\begin{minipage}{0.48\textwidth}
 \includegraphics[width=\textwidth,height=5cm]{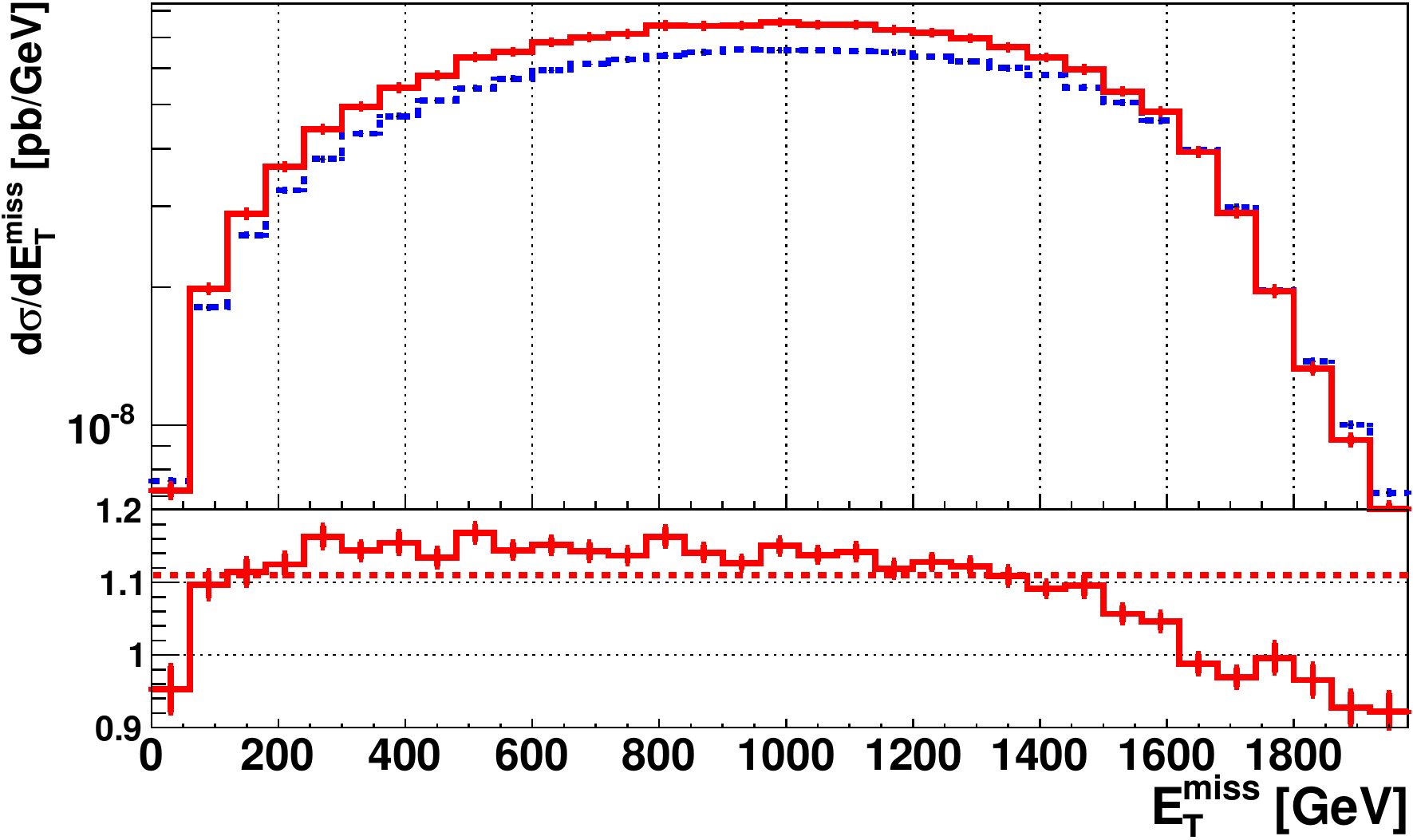}
 \vspace{0.1cm}
 \newline
 \includegraphics[width=\textwidth,height=5cm]{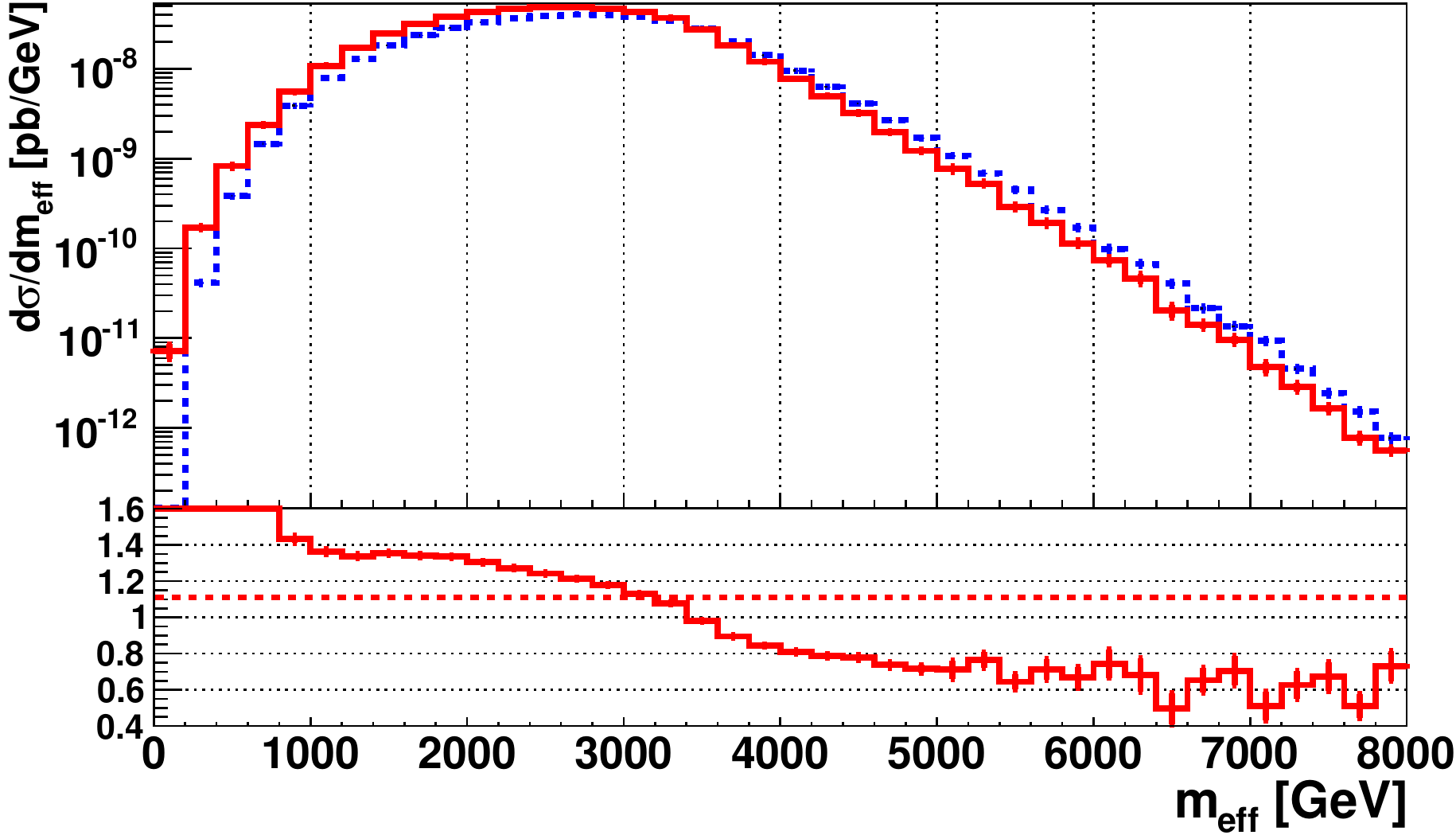}
\end{minipage}
\caption {\label{fig:LONLOdec1} Differential distributions as defined in the text for squark-antisquark production, combined with the subsequent decay $\sq\rightarrow q \neutone$ and the corresponding decay for the antisquark for the scenario $10.3.6^*$. Shown are the LO predictions obtained using Eq.~(\ref{eq:LOproddec}) and the NLO results determined according to Eq.~(\ref{eq:proddec1}). In all plots the lower panels depict the respective differential $K$-factor (full) and the total $K$-factor from Tab.~\ref{tab:proddec} (dashed).}
\efig
Next we consider the same set of distributions for squark pair production with subsequent decays, this time for the scenario $10.4.5$, depicted in Fig.~\ref{fig:LONLOdec3}. 
Again the results for $10.3.6^*$ are qualitatively identical and not shown here. In essence, the behaviour is very much the same as for squark-antisquark production in Fig.~\ref{fig:LONLOdec1} and differs only in details. For example the differential $K$-factor of the rapidity distribution $y^{j_1}$ shows slightly larger deviations from the total $K$-factor, whereas the one for $\slashed{E}_T$ is a bit flatter in the range considered here.
\bfig[t]
  \begin{minipage}{0.48\textwidth}
 \includegraphics[width=\textwidth,height=5cm]{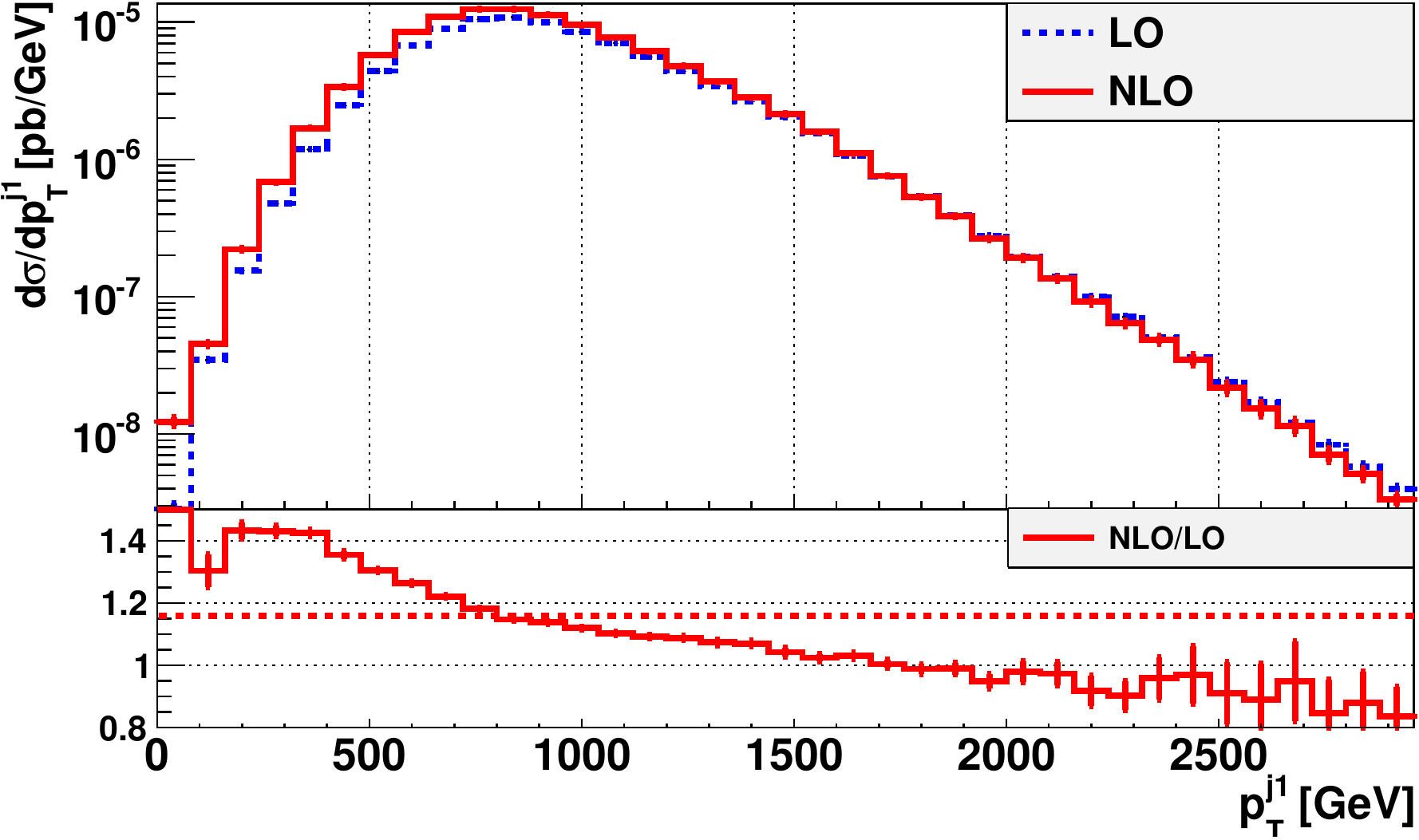}
 \vspace{0.1cm}
 \newline
 \includegraphics[width=\textwidth,height=5cm]{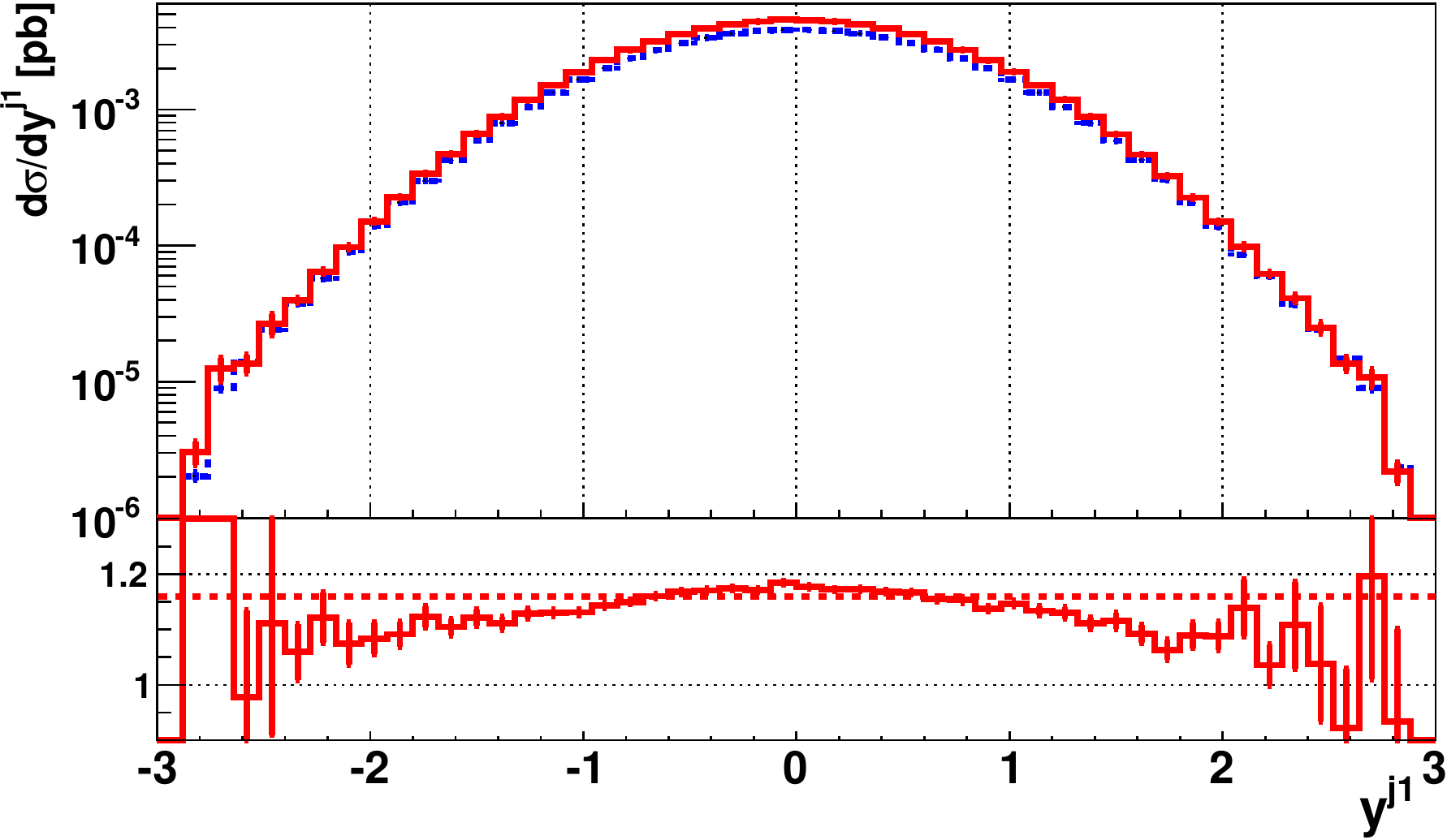}
\end{minipage}
\begin{minipage}{0.04\textwidth}
\end{minipage}
\begin{minipage}{0.48\textwidth}
 \includegraphics[width=\textwidth,height=5cm]{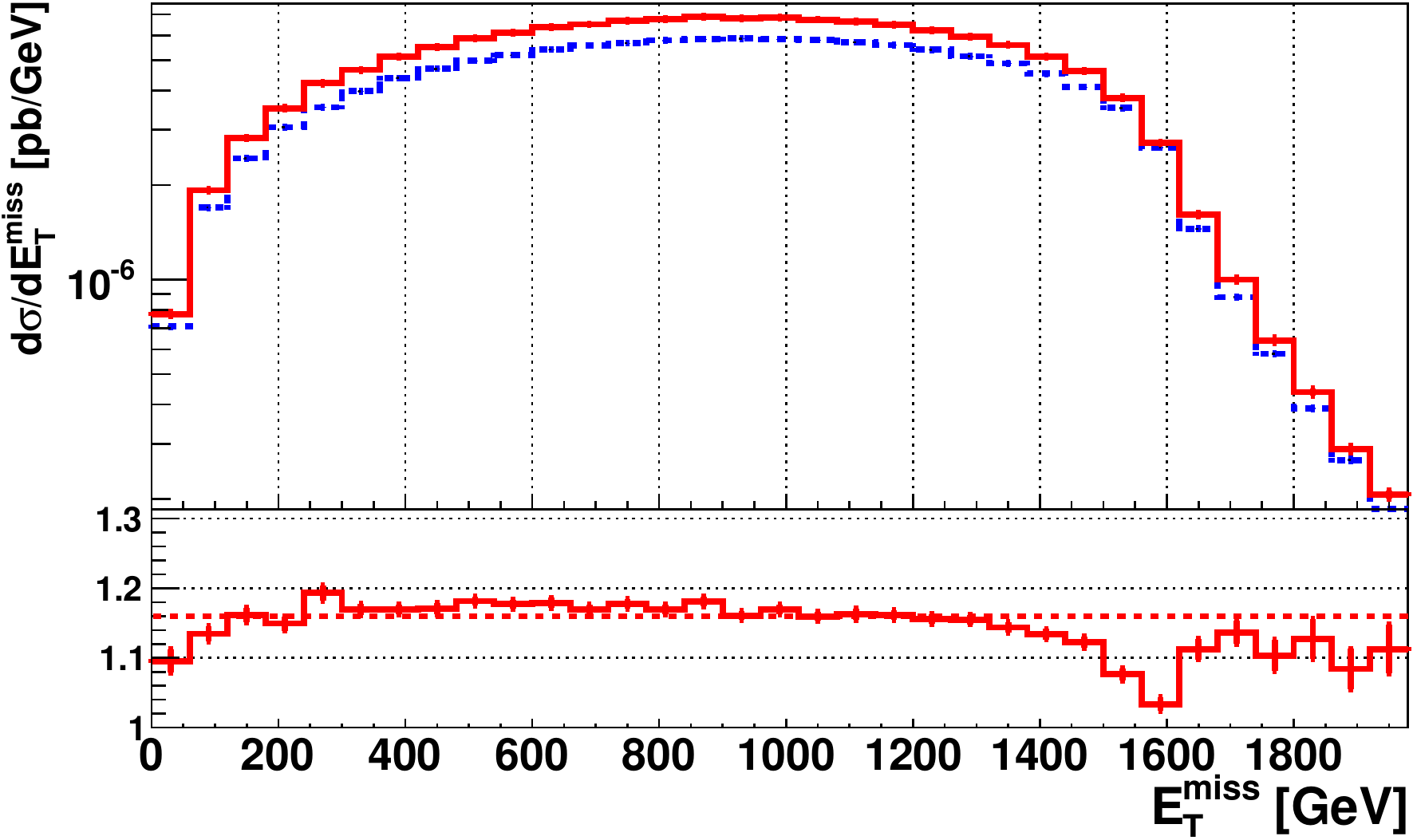}
 \vspace{0.1cm}
 \newline
 \includegraphics[width=\textwidth,height=5cm]{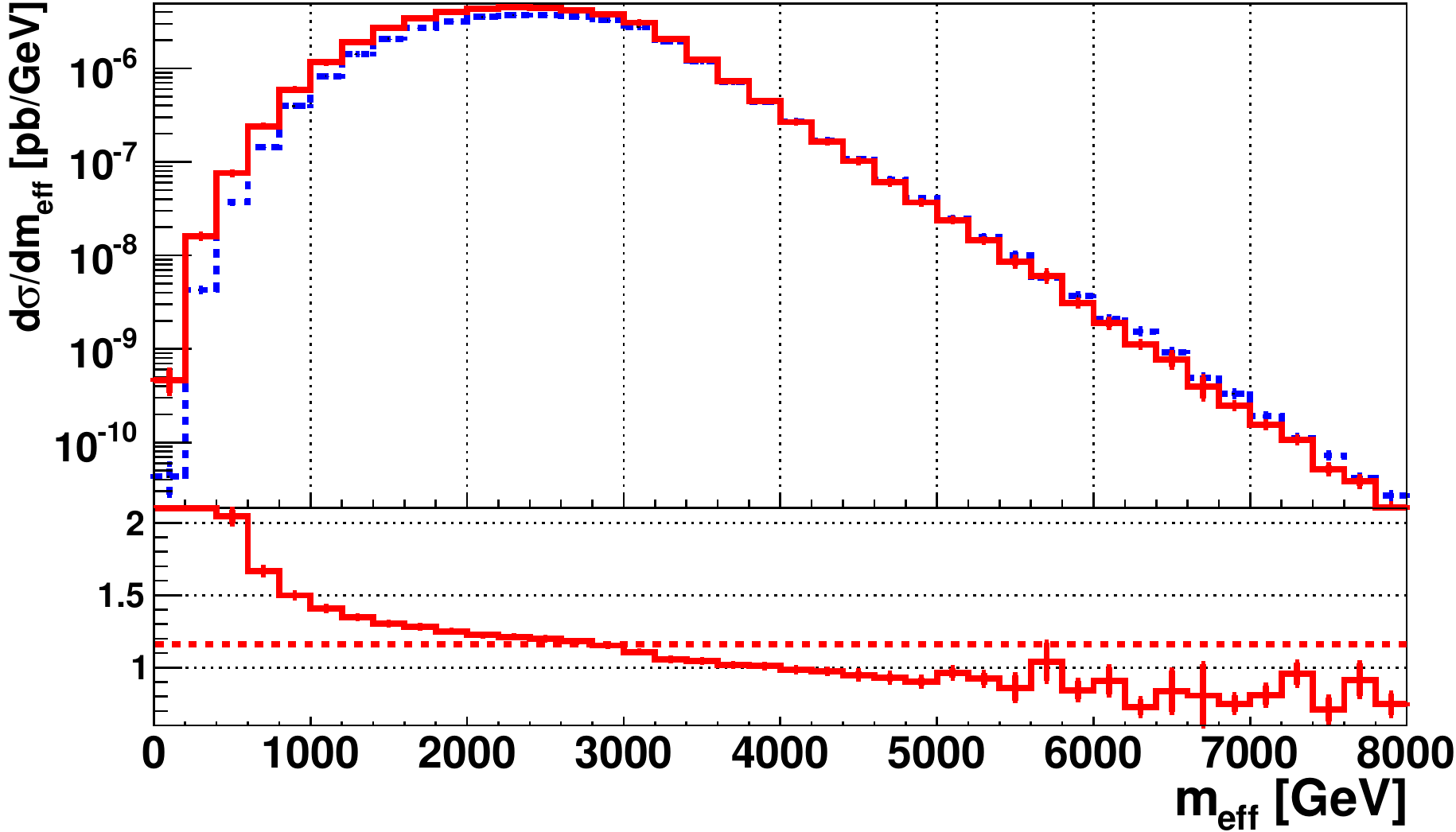}
\end{minipage}
\caption {\label{fig:LONLOdec3} Same as Fig.~\ref{fig:LONLOdec1} for squark pair production and the scenario $10.4.5$.}
\efig

\subsubsection{Parton Shower Effects}
In order to investigate parton shower effects we have combined our implementations of the squark production and decay processes with different parton shower programs. To this end five million events have been generated for squark-antisquark and squark pair production for each of the two benchmark scenarios defined in Sec.~\ref{sec:setup}. 
The results shown in the following have been obtained by setting the folding parameters of the \PB~to the values
\be
 n_{\xi}=5,\quad n_y=5,\quad n_{\phi}=1\,,
\ee
reducing such the number of events with negative weights.
However, in the context of squark production and decay processes two further sources of negative events can occur. The first one originates from the way production and decay are combined in Eq.~(\ref{eq:proddec1}), see the discussion in Sec.~\ref{sec:proddec}. It is not possible to apply the folding procedure described above in this case, since the negative contributions to $\overline\matB$ are directly related to the (modified) Born contribution. Using a different expansion of the combination formula, {\it e.g.} Eq.~(\ref{eq:proddec2}), would remedy this point, however this approach violates unitarity and should therefore be avoided. 

The implemented subtraction schemes described in Sec.~\ref{sec:virtreal} present another source of contributions with negative weights. While these are completely absent for the DR-I method and their number can be reduced again by means of folding for the DS$^*$ and the DR-II method, they inevitably occur for the methods relying on a splitting of $\matR$.

All in all, using the (for conceptual reasons preferable) DS$^*$ subtraction scheme with split real matrix elements squared and Eq.~(\ref{eq:proddec1}) for the combination of production and decay leads unavoidably to events with negative weights, which cannot be neglected. Therefore, they are kept in the generated event files by setting the \PB~flag {\tt withnegweights=1}. 

The generated event samples have been showered with two Monte Carlo event generators, using three different parton shower algorithms implemented in these programs:
\bi
\item \textbf{\textsc{Pythia 6}:} We use the version 6.4.28 \cite{pythia6}. All results have been obtained with the Perugia 0 tune \cite{perugia}, invoked by setting {\tt MSTP(5) = 320}. A comparison with the Perugia 11 tune ({\tt MSTP(5) = 350}) yields only tiny discrepancies.\footnote{To be more precise, for squark-antisquark production, including the decays and using the benchmark scenario $10.3.6^*$, of all observables considered in this section only the $p_T^{j_3}$ distribution shows with $\matO(5\%)$ a deviation larger than 1\%.} In order to study only effects of the parton shower, hadronization and multi-parton interaction (MPI) effects have been turned off by setting {\tt MSTP(111) = 0} and {\tt MSTP(81) = 20}, invoking thus the use of the $p_T$-ordered shower. 

However, in the simulation of the full process, including NLO corrections to the production and the decays, a further subtle difficulty arises when using \textsc{Pythia}, which is related to the way the starting scales for the shower are chosen. The \textsc{Powheg} approach relies on the assumption that the $p_T$ of the emitted final-state parton is larger than the transverse momentum of any subsequent splitting generated by the parton shower. This requires the application of a $p_T$ veto in the parton shower, with the maximal scale being read for each event from the event file. However, if final-state resonances are present the mass of those has to be preserved by the reshuffling operations performed in the shower algorithm. Therefore, the showering of partons originating from the decays of these resonances, {\it i.e.} in the processes considered here the produced squarks, is performed separately in \PYTH. The starting scale for these shower contributions is set to the invariant mass of all decay products, hence in the case at hand to the mass of the respective squark. In the scenarios considered here this scale is typically an order of magnitude larger than the upper scale written to the event file, leading to much more radiation and thus to a strong bias of the results. In order to correct for this effect, the \textsc{Pythia} routines had to be adapted to use the scale specified in the event file as starting scale in all individual contributions to the parton shower. 

\item \textbf{\textsc{Herwig++}:} The default shower of \textsc{Herwig++} \cite{herwigpp} is ordered in the angles of the branchings. Applying this shower to an event sample generated according to the \textsc{Powheg} method requires again the use of a $p_T$ veto. However, this combination lacks the emission of soft wide-angle partons, as the first emission in an angular-ordered shower is not necessarily the hardest one. In principle these missing parts have to be simulated in an extra step via a so-called vetoed truncated shower, which is not provided by \textsc{Herwig++} and thus not taken into account in the following. The effect of this missing part will be estimated by comparing the results to those obtained with the $p_T$-ordered \textsc{Dipole-Shower} \cite{herwigdp1,herwigdp2}, which is also part of the \textsc{Herwig++} framework. The results presented in the following sections have been obtained using the version 2.6.1 \cite{herwigpp26}. 
 In the following \HWG~refers only to the default shower, while the results labeled \DS~or, for the sake of brevity, \textsc{Dipole}~refer to the \DS~included in the \HWG~framework. 
\ei

The showered results for squark-antisquark production are shown in Fig.~\ref{fig:Shower1}, using the scenario $10.3.6^*$.
Likewise, Fig.~\ref{fig:Shower2} depicts the results for squark pair production, obtained with the scenario $10.4.5$. All plots show the outcome of the three parton showers described above and the NLO prediction, which serves as normalization in the ratio plots shown in the lower panels. The results for squark pair production using scenario $10.3.6^*$ and squark-antisquark production with scenario $10.4.5$ do not reveal any new features compared to the depicted combinations and are therefore not shown here.
\bfig[t]
  \begin{minipage}{0.48\textwidth}
 \includegraphics[width=\textwidth,height=5cm]{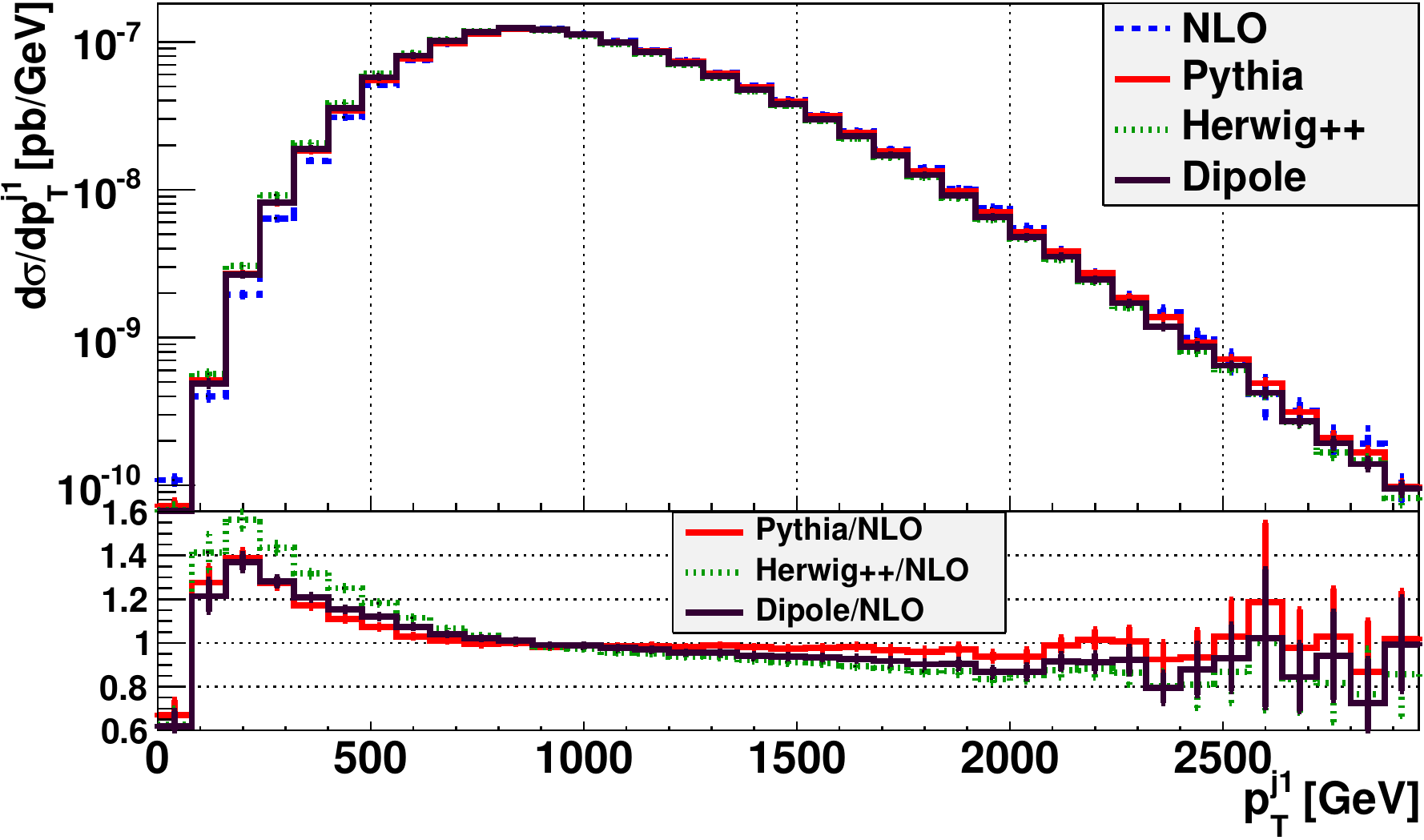}
 \vspace{0.1cm}
 \newline
 \includegraphics[width=\textwidth,height=5cm]{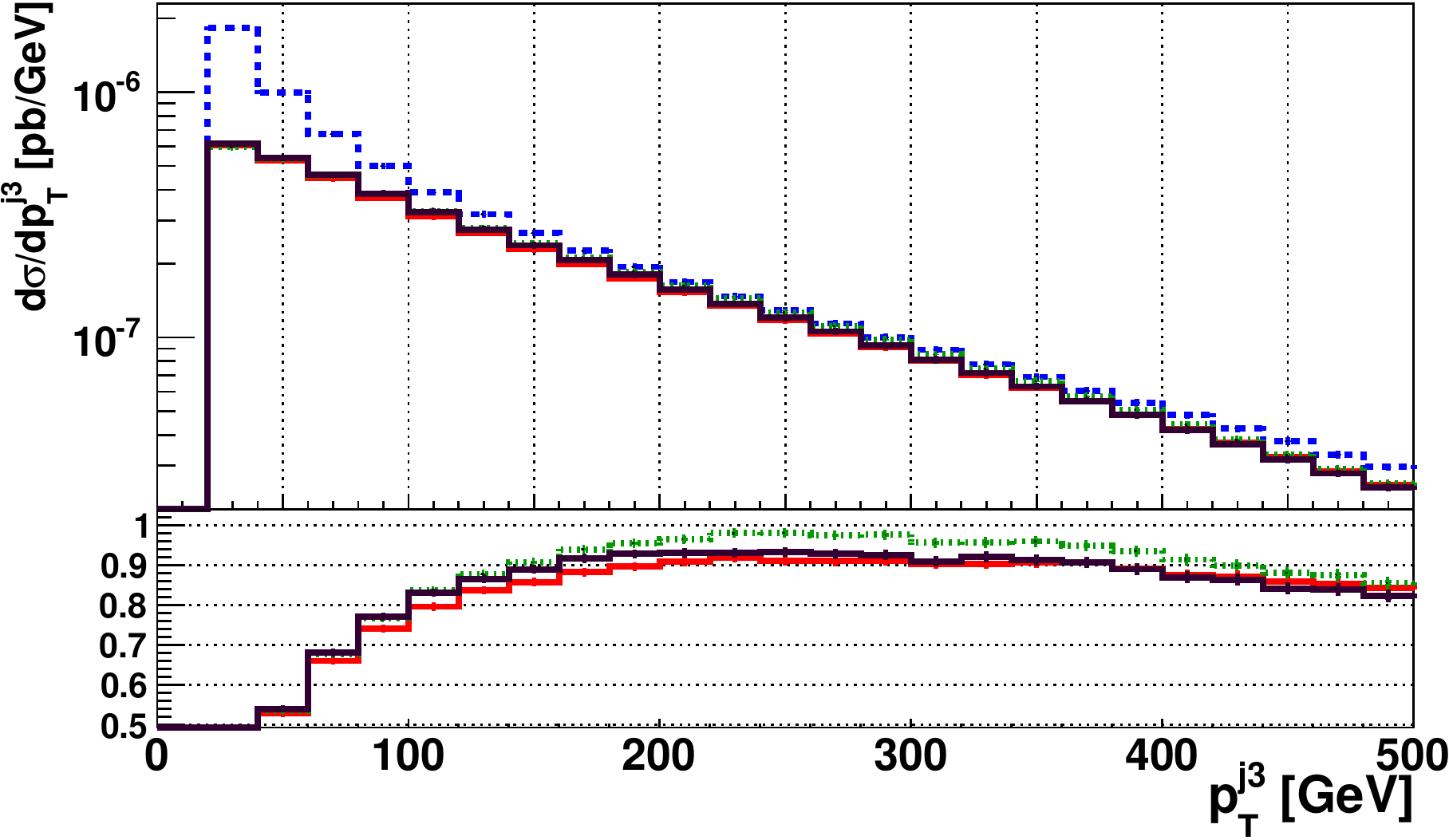}
\end{minipage}
\begin{minipage}{0.04\textwidth}
\end{minipage}
\begin{minipage}{0.48\textwidth}
 \includegraphics[width=\textwidth,height=5cm]{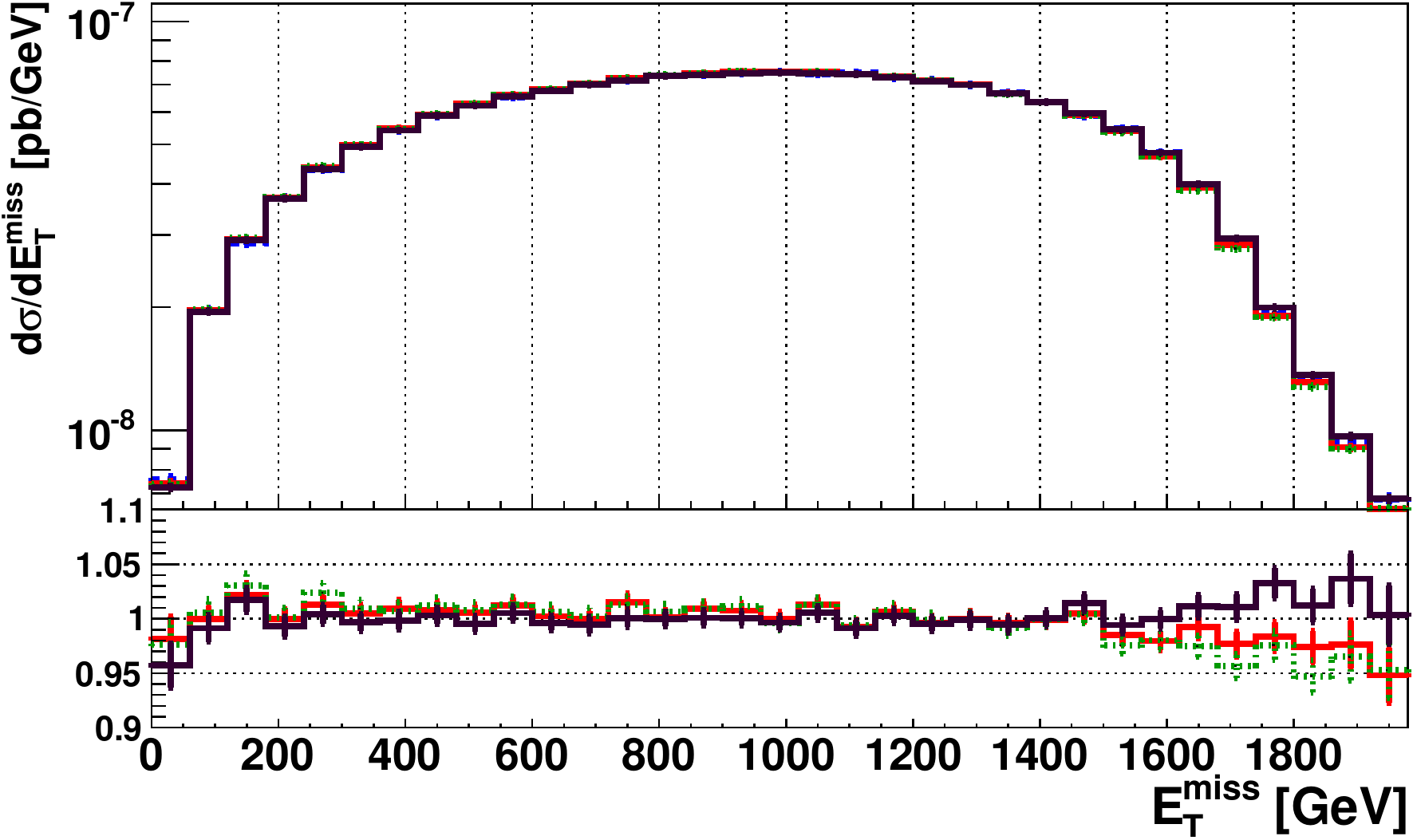}
 \vspace{0.1cm}
 \newline
 \includegraphics[width=\textwidth,height=5cm]{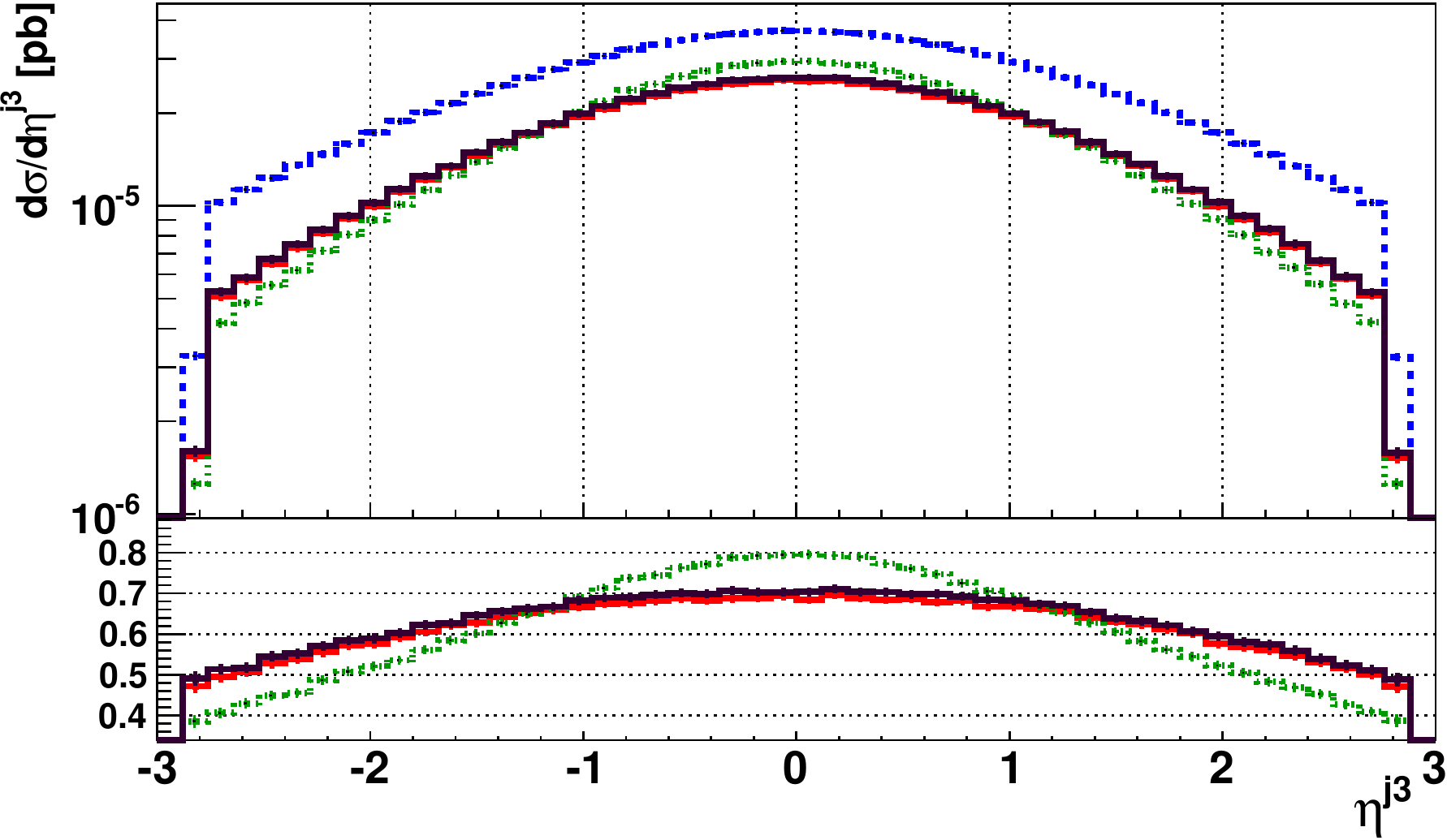}
\end{minipage}
\caption {\label{fig:Shower1} Differential distributions for squark-antisquark production, combined with the subsequent decay $\sq\rightarrow q \neutone$ for the scenario $10.3.6^*$. The NLO predictions and the results after applying the parton showers \PYTH, \HWG and the \DS\ are shown. In all plots the lower panels depict the respective ratios of the results obtained with the three parton showers and the pure NLO prediction.}
\efig

Comparing the predictions for the individual observables shown in the two figures we note that in all cases considered here the $p_T^{j_1}$ result obtained with \HWG~is in the low-$p_T$ region slightly ($\matO(10\%)$) enhanced compared to the other parton showers, whereas the \DS~and \PYTH~essentially agree here. At the other end of the spectrum, however, both the \HWG~and the \DS~predict $\matO(10\%)$ smaller rates than \PYTH, which is almost in accordance with the NLO result for large values of $p_T^{j_1}$. 
The outcome of \HWG~and the \DS~is identical in this kinematic regime. Similar conclusions can be drawn from the $p_T^{j_2}$, the $m^{j_1j_2}$ and the $m_{\text{eff}}$ distributions not shown here. In contrast, the distributions describing the third hardest jet show more pronounced differences.
Comparing first the results for $p_T^{j_3}$ obtained with \HWG~and the \DS~one notices that they agree within $\matO(5-10\%)$. The result for the third jet obtained with \PYTH~is in all cases smaller compared to the other two parton showers. While the discrepancy using the benchmark scenario $10.3.6^*$ is for both squark-antisquark and squark pair production smaller than 10\%, it amounts to 10-15\% for the scenario $10.4.5$ in both cases.
The largest differences in the three shower predictions emerge in the results for the pseudorapidity of the third hardest jet, $\eta^{j_3}$. While \PYTH~and the \DS~agree within 5\% for all cases and differ in case of squark pair production only in the overall normalization, but not in the shape of the distributions, \HWG~predicts evidently more jets in the central region $\left|\eta^{j_3}\right|\lesssim 1$. Comparing the \HWG~result and the \PYTH~outcome for squark-antisquark production, this enhancement amounts to a 20\% higher rate in the centre and a reduction of the same magnitude for $\left|\eta^{j_3}\right|\approx 2.8$. In case of squark pair production, this effect is smaller, of $\matO(10\%)$, but still clearly visible. 
The predictions for the missing transverse energy $\slashed{E}_T$ agree very well and essentially reproduce the NLO result. Tiny deviations are only visible in the tails of the distributions, however they are smaller than 5\% in all cases.

\bfig[t]
  \begin{minipage}{0.48\textwidth}
 \includegraphics[width=\textwidth,height=5cm]{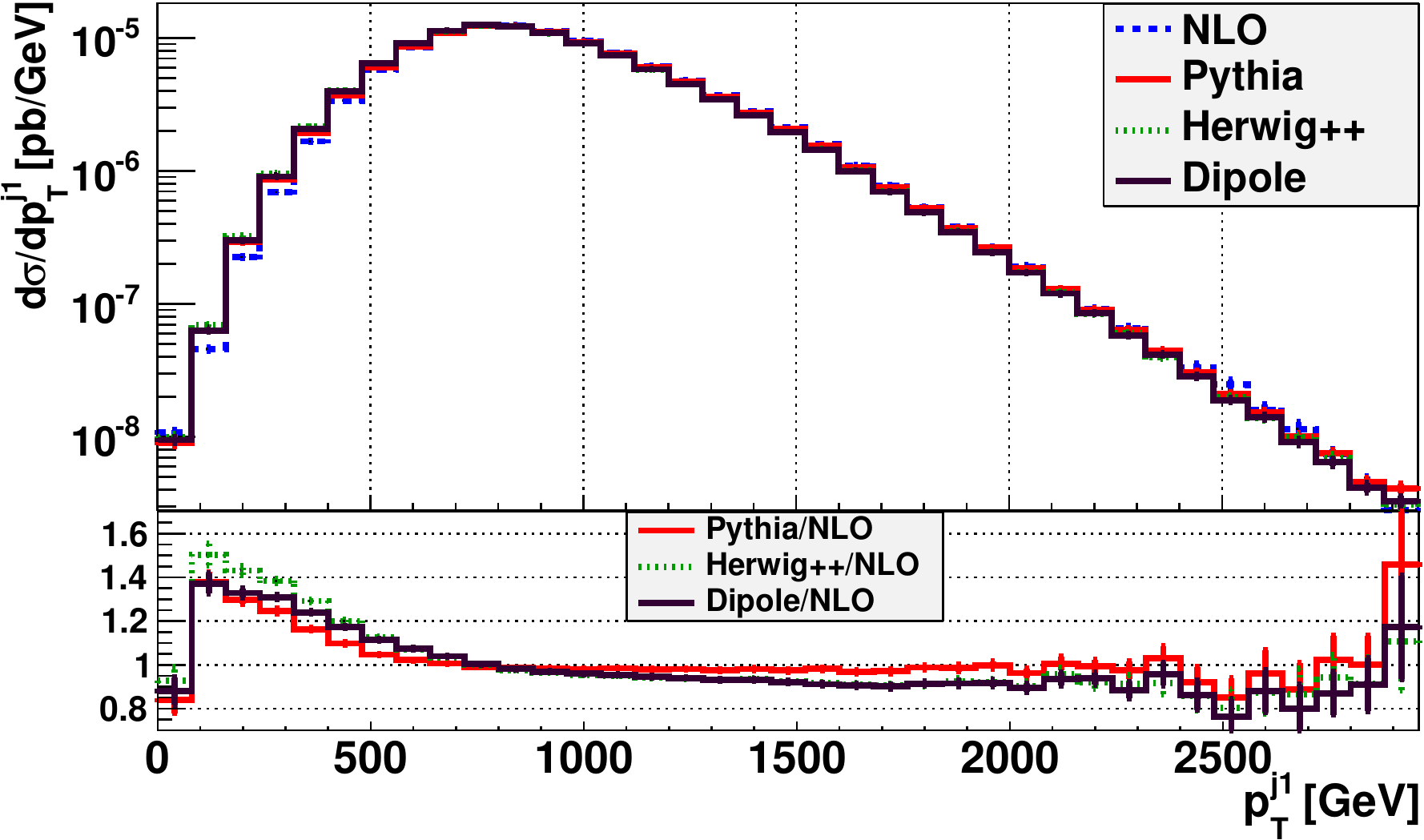}
 \vspace{0.1cm}
 \newline
 \includegraphics[width=\textwidth,height=5cm]{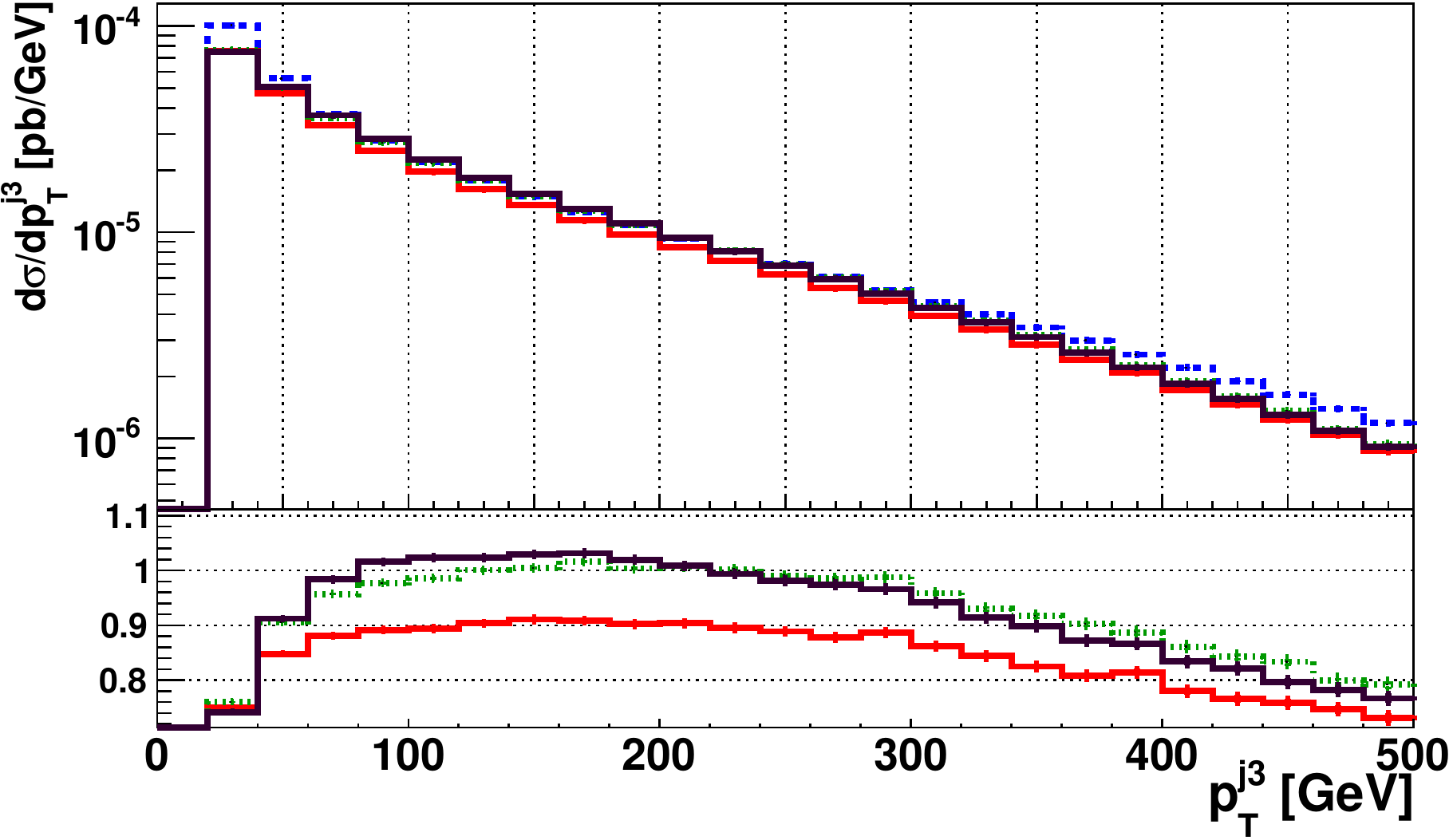}
\end{minipage}
\begin{minipage}{0.04\textwidth}
\end{minipage}
\begin{minipage}{0.48\textwidth}
 \includegraphics[width=\textwidth,height=5cm]{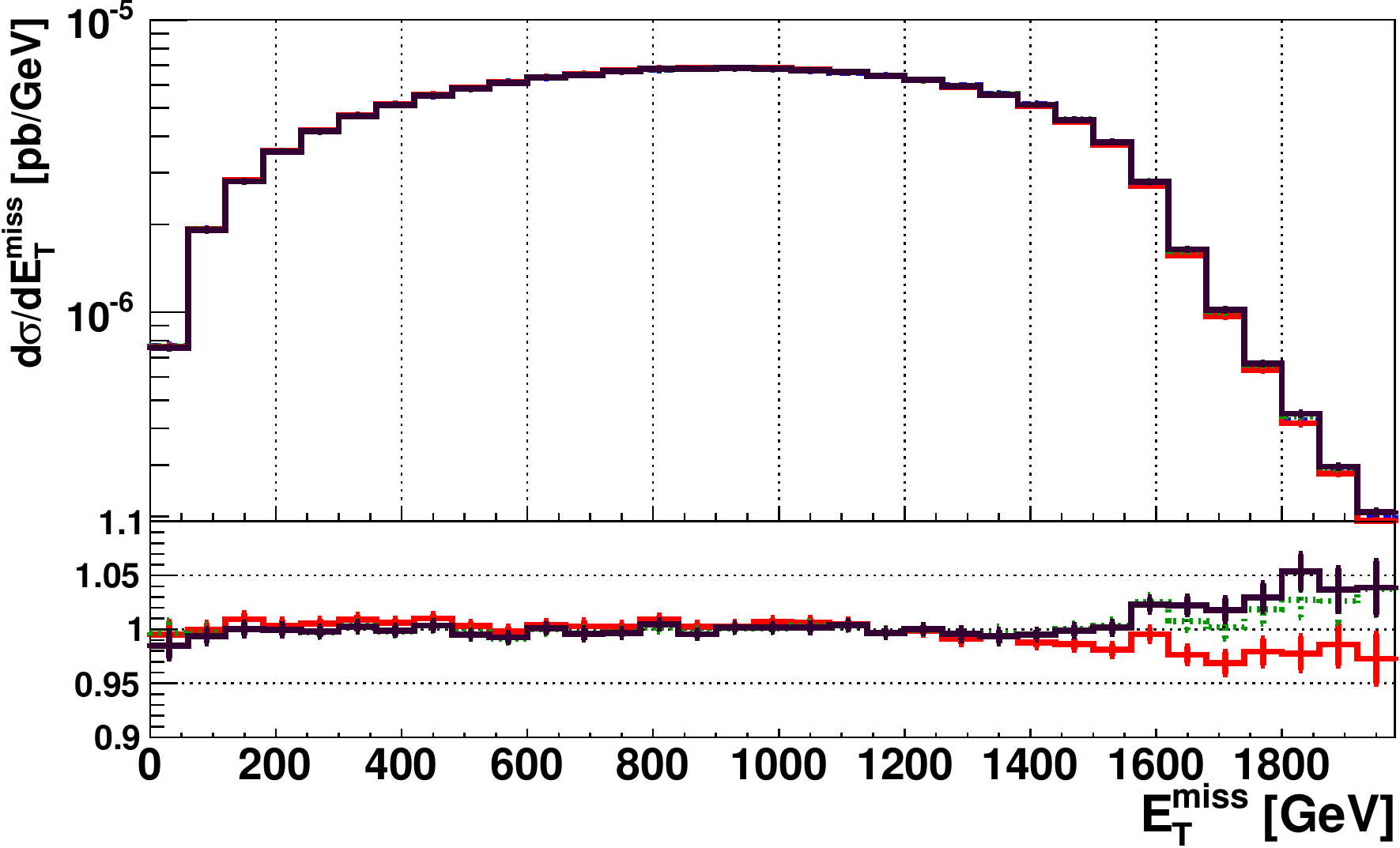}
 \vspace{0.1cm}
 \newline
 \includegraphics[width=\textwidth,height=5cm]{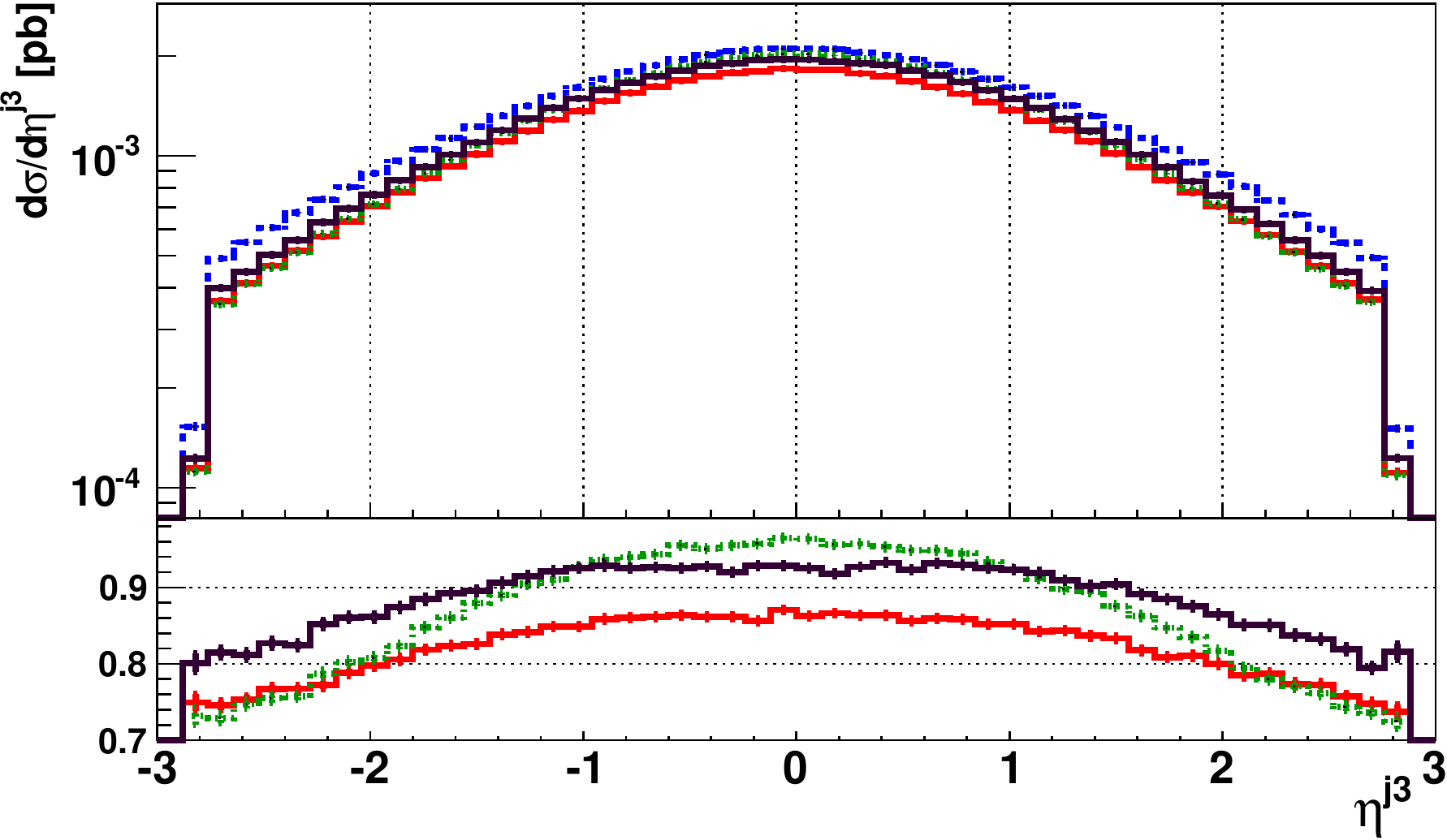}
\end{minipage}
\caption {\label{fig:Shower2} Same as Fig.~\ref{fig:Shower1} for squark pair production and the benchmark scenario 10.4.5.}
\efig

All in all, the predictions of the different parton showers for the observables depending solely on the two hardest jets agree within $\matO(10\%)$ or better. Comparing the showered results with the outcome of a pure NLO simulation the effects of the parton showers on these observables are at most of $\matO(10-20\%)$, except for the threshold region. By and large, the two \HWG~showers yield larger deviations from the NLO outcome for these observables, whereas \PYTH~reproduces the NLO curves within $\matO(10\%)$. The $\slashed{E}_T$ distribution is in all cases hardly affected by parton shower effects. 
Larger deviations between the different parton showers emerge in the predictions for the third hardest jet, which is formally described only at LO in the hard process. Especially the \HWG~prediction differs significantly from the other two showers and predicts more jets in the central region of the detector. At this point it is not possible to decide ultimately if this discrepancy is an effect of the missing truncated shower or simply a relict of the way the phase space is populated in the different shower algorithms. This would require the actual implementation of a vetoed truncated shower, which is beyond the scope of this work.
However, comparing the outcomes of the \DS~and \HWG~reveals only very small discrepancies in other observables. Hence the overall effect of the neglected truncated shower seems to be small. 
\subsubsection{Total Rates}
\label{sec:totrat}
In the last step the created event samples are analysed using a realistic set of event selection cuts, which corresponds to the definition of the signal region \lq A-loose' for the SUSY searches in two-jet events performed by the ATLAS collaboration \cite{atlasexcl4}.

The event selection cuts used in this analysis are
\begin{gather}
  p_T^{j_1}>130\,\text{GeV}, \quad p_T^{j_2}>60\,\text{GeV},  \quad \slashed{E}_T>160\,\text{GeV},\quad  \frac{\slashed{E}_T}{m_{\text{eff}}}>0.2,\quad m_{\text{eff}}^{\text{incl}}>1\,\text{TeV},\nonumber\\
  \Delta\phi(j_{1/2},\vec{\slashed{E}}_T)>0.4,\quad \Delta\phi(j_3,\vec{\slashed{E}}_T)>0.4 \,\,\,\,\,\text{if}\,\,\,\,\, p_T^{j_3}>40\,\text{GeV}\,.
\label{eq:atlascuts}
\end{gather}
Here, the effective mass $m_{\text{eff}}$ is defined as the sum of the $p_T$ of the two hardest jets and $\slashed{E}_T$, whereas the inclusive definition of this observable includes all jets with $p_T^j>40\,\text{GeV}$,
\be
 m_{\text{eff}}^{\text{incl}} = \sum_{i=1}^{n_j} p_T^{j_i} + \slashed{E}_T\,.
\ee
Moreover, $\Delta\phi(j_i,\vec{\slashed{E}}_T)$ denotes the minimal azimuthal separation between the direction of the missing transverse energy, $\vec{\slashed{E}}_T$, and the $i^{\text{th}}$ jet. The additional cut $\Delta\phi(j_3,\vec{\slashed{E}}_T)>0.4$ is only applied if a third jet with $p_T^{j_3}>40\,\text{GeV}$ is present.

\begin{table}
\renewcommand{\arraystretch}{1.2}
\bc
\begin{tabular}{|c || c |c  || c | c | }\hline
      & \multicolumn{2}{c||}{$10.3.6^*$} & \multicolumn{2}{c|}{$10.4.5$} \\\cline{2-5}
      & $\sq\sq$ & $\sq\sqbar$ & $\sq\sq$ & $\sq\sqbar$\\\hline\hline
NLO   & $0.871\,\text{fb}$  & $0.0781\,\text{fb}$  &  $6.809\,\text{fb}$  & $0.696\,\text{fb}$ \\  
\PYTH & $0.883\,\text{fb}$  & $0.0797\,\text{fb}$  & $6.854\,\text{fb}$ & $0.704\,\text{fb}$ \\
\HWG  & $0.895\,\text{fb}$  & $0.0807\,\text{fb}$  & $6.936\,\text{fb}$ & $0.711\,\text{fb}$ \\\hline
\PYTH~(approx.) & $0.855\,\text{fb}$  & $0.0664\,\text{fb}$  & $6.844\,\text{fb}$ & $0.617\,\text{fb}$ \\
\HWG~(approx.) & $0.858\,\text{fb}$  & $0.0667\,\text{fb}$  & $6.876\,\text{fb}$ & $0.620\,\text{fb}$ \\\hline
\end{tabular}
  \caption{\label{tab:sigcuts} Total cross sections after applying the event selection cuts defined in Eq.~(\ref{eq:atlascuts}) for the different production modes in the two benchmark scenarios. The decays of the squarks (antisquarks) to $q\neutone$ ($\bar{q}\neutone$) are included at NLO. The given results have been obtained at the level of a pure NLO simulation and including parton shower effects with \PYTH~and \HWG, respectively. The last two rows have been obtained by rescaling LO events after application of the \PYTH/\HWG~shower with the constant $K$-factor and the individual NLO branching ratios.}
\ec
\vspace*{-0.2cm}
\end{table}

Applying these cuts at the level of a pure NLO simulation yields for the total cross sections for squark (anti)squark production combined with the subsequent decays in the two benchmark scenarios $10.3.6^*$ and $10.4.5$ the results given in the first row of Tab.~\ref{tab:sigcuts}. Matching these NLO results with a parton shower hardly affects the outcome after using the cuts defined in Eq.~(\ref{eq:atlascuts}), as can be inferred from the results obtained with \PYTH~and the \HWG~default shower listed in the second and third row, respectively. Note that due to the mixture of cuts on inclusive and exclusive quantities the rates predicted by the two showers are slightly larger compared to the NLO case. Moreover, the two parton showers yield identical rates within 1-2\%.

In order to compare these results obtained with our new calculations and implementations with the values determined according to the setup used so far for the simulation of these processes we proceed as follows: first the production and decays of the squarks are simulated with LO accuracy. The resulting events are reweighted with a common $K$-factor for squark-antisquark or squark pair production, which is obtained from \textsc{Prospino}, {\it i.e.} assuming degenerate squark masses and averaging over all channels. Each individual production channel is then multiplied with the corresponding NLO branching ratios for the produced squarks. The rescaled events are subsequently processed with the \PYTH~and the \HWG~default shower, neglecting again effects of hadronization, MPI, etc. 
The results obtained with this approximate setup after applying the event selection cuts defined in Eq.~(\ref{eq:atlascuts}) are summarized in the last two rows of Tab.~\ref{tab:sigcuts}. Comparing these total rates with those obtained in the full simulation one notes that the discrepancy is almost negligible in case of squark pair production, but amounts to 15-20\% for squark-antisquark production. This discrepancy is mainly caused by assuming a common $K$-factor for all subchannels instead of using the exact results with individual $K$-factors when combining production with decay. This effect in squark-antisquark production has already been demonstrated in section \ref{sec:fixedorder} for the case of LO decays. In squark pair production, however, subchannels with $K$-factors close to the global $K$-factor have large branching ratios and therefore the exact and the approximate method give similar results. These examples illustrate that in order to obtain precise predictions it is not in all cases sufficient to use the approximate approach.

\section{Conclusions}
\label{ch:conclusion}
One of the main tasks of the LHC is the search for beyond the SM
physics, in particular supersymmetry. At the high-energy run of the
LHC coloured SUSY particles can be produced with masses up to the
multi-TeV range. In order to find these particles and be able to
measure their properties, reliable predictions for the production cross
sections both at the inclusive and at the exclusive level are
mandatory. In this paper we continue our effort in providing
accurate theoretical predictions by presenting results for the
squark-antisquark production of the first two generations at NLO
SUSY-QCD without making any simplifying assumptions on the sparticle masses and by
treating the different subchannels individually. As developed in our
previous paper \cite{ownpaper} we have performed the subtraction of
possible on-shell intermediate gluinos in a gauge-invariant
approach and compared to several methods proposed in the
literature. While in squark pair production for the investigated
scenarios the differences in the 
total rates turned out to be negligible  and quite small for
distributions, in squark-antisquark production, where the contributions
of the $qg$ initiated channels are more important, larger differences
were found. They amount to about 4\% for the inclusive NLO cross section in the
investigated scenario. Even larger effects are found in the
distributions, where the discrepancies between the investigated
methods can be up to 30\%  in the $p_T$ distribution of the radiated
parton. The invariant 
mass distribution of the squark-antisquark pair is not affected by the
chosen method, however, and only reflects the discrepancy in the total
cross section. 

The $K$-factor for squark-antisquark production has been found to be 
sizeable and positive with $K \equiv \sigma_{\rm NLO}/\sigma_{\rm LO}\approx 1.4$ and the scale uncertainty is
strongly reduced by taking into account the NLO corrections. The
comparison of the results for individual $K$-factors
for the subchannels contributing to squark-antisquark production  and
the $K$-factor obtained after summing the cross sections differ
significantly, so that the use of a global $K$-factor in general does
not lead to accurate predictions. Combining the NLO production cross
section with LO decays of the (anti)squark into the lightest
neutralino and (anti)quark leads to discrepancies of about 10\% between the
exact result and the one assuming a common $K$-factor. The more the
branching ratios of the squarks for the specific decay channel under
consideration differ, the more important the consistent treatment of
the individual corrections becomes. 

In a next step we have combined our results for squark pair and
for squark-antisquark production with the decays of the final state
(anti)squarks into the lightest neutralino and (anti)quark at NLO
SUSY-QCD at fully differential level. In this context we have
discussed two methods for the combination of production and decay
with NLO accuracy in the kinematics. One is based on a Taylor expansion
respecting unitarity, but suffering from possibly negative
contributions. The second approach, which does not expand the total
decay width entering the branching ratios of the decays, avoids this
problem, however violates unitarity. The results for these two
approximations and for the case where no expansion in the strong 
coupling is performed at all, differ by at most 4\% for the
total cross sections. In the jet distributions the discrepancies
between the two approximations can be up to 15\%, whereas in
the $\slash{\!\!\!\! E}_T$ distribution they are purely given by the
discrepancy in the total cross sections. In view of these small
deviations, in particular for the inclusive quantities, we have adopted
the unitarity preserving approach in the remaining numerical analysis. 

The influence of the NLO corrections on the distributions has been
investigated for several observables. While in the $\slash{\!\!\!\! E}_T$
distribution the deviation of the differential $K$-factor and the
total one is of ${\cal O}(5\%)$, the $K$-factor for the $p_T$ distributions of the two
hardest jets can vary in a range of $\pm 40$\%, hence the assumption
of a constant $K$-factor clearly is not valid here any more. 

In order to obtain realistic predictions for exclusive observables we
have matched the NLO cross sections with parton showers using
the \textsc{Powheg-Box}  framework. The
implementation is publicly available and can be downloaded from 
\cite{powhegweb}.

The matched NLO results have been interfaced with the $p_T$ ordered
shower of \textsc{Pythia6} as well as the default shower and the
Dipole shower of \textsc{Herwig++}. To allow for a consistent comparison
of the three showers, in \textsc{Pythia} the starting scale for the radiation off the
decay products had to be modified. The largest differences in the
three shower predictions is found in the pseudorapidity distribution of the third
hardest jet. Thus \textsc{Herwig++} predicts more jets in the central
region, which is particularly pronounced for squark-antisquark
production in the investigated scenario. The comparison of the
showered result with a pure NLO simulation shows small differences for
more inclusive quantities. In more exclusive distributions, in
particular \textsc{Herwig++} shows large deviations from the pure NLO
result, as {\it e.g.}\ in the predictions for the third hardest jet. To
decide if this is an effect of the missing truncated shower or a
relict of the way the phase space is populated would require further
investigations and is beyond the scope of this work. 

Finally, we performed a cut-based analysis of the total
cross sections in two benchmark scenarios using realistic event
selection cuts taken from an ATLAS analysis. Comparing our results with
the approximate approach used by the experiments revealed small
discrepancies for squark pair production, but up to 20\% differences
for squark-antisquark production. This effect could be traced back to 
assuming a common $K$-factor for the production cross sections of all 
subchannels instead of using the exact results. These examples show
that the effects can be sizeable and precise theoretical predictions
should take into account the full NLO calculation for the production
processes, consistently combined with the squark decays at NLO. 

The reliable exploitation and interpretation of the LHC data in the
search for new physics requires accurate theoretical predictions
for production and decay of SUSY particles including higher order
corrections not only for 
inclusive quantities but also for distributions. Our results for the
fully differential calculation of the SUSY-QCD corrections to squark
pair and squark-antisquark production combined with their subsequent
decay at NLO SUSY-QCD and matched with parton showers show 
that the independent treatment of the contributing
subchannels is essential and that differential $K$-factors can not be assumed to
be flat. The results presented here are the next step in our program
of providing a fully differential description of SUSY particle
production and decay at the LHC. 
\section*{Acknowledgments}

M.P. thanks Laura Jenniches and Alexander M\"uck for useful discussions. 
C.H. has been supported by the \lq Karlsruher Schule für Elementarteilchen- und Astroteilchenphysik: Wissenschaft und Technologie (KSETA)'. 
This work has been supported by the DFG SFB/TR9 “Computational Particle Physics”.
The research of M.S. is supported in part by the European Commission through the “HiggsTools” Initial Training Network PITN-GA-2012-316704.  
Last but not least, we thank Carlo Oleari for making our code publicly available via the \PB\ website.  



\bibliographystyle{utphys.bst}
\bibliography{paper}
\end{document}